\documentclass[12pt,preprint]{aastex}
\usepackage{natbib}

\def\eg{\mbox{e.g.}}
\def\ie{\mbox{i.e.}}
\def\kms{\mbox{km~s$^{\rm -1}$}}

\newcommand{\HI}{H\,{\sc i}}

\slugcomment{Resubmitted to ApJ}

\shorttitle{GASS High-Velocity Clouds in the Region of the Magellanic Leading Arm}

\shortauthors{B.-Q. For, L. Staveley-Smith, N.M. McClure-Griffiths}

\begin{document}

\title{GASS High Velocity Clouds in the Region of the Magellanic Leading Arm}

\author{Bi-Qing For\altaffilmark{1}, Lister Staveley-Smith\altaffilmark{1}, 
N.\ M.\ McClure-Griffiths\altaffilmark{2}}

\affil{$^1$International Centre for Radio Astronomy Research, University of Western Australia, 
35 Stirling Hwy, Crawley, WA, 6009, Australia; biqing.for@uwa.edu.au}
\affil{$^2$CSIRO Astronomy and Space Science, Epping, NSW, 1710, Australia} 

\begin{abstract}
We present a catalog of high-velocity clouds in the region of the Magellanic Leading Arm. 
The catalog is based on neutral hydrogen (\HI) observations 
from the Parkes Galactic All-Sky Survey (GASS).
Excellent spectral resolution allows clouds with narrow-line components to be resolved. 
The total number of detected clouds is 419. 
We describe the method of cataloging and present 
the basic parameters of the clouds.
We discuss the general distribution of the high-velocity clouds and 
classify the clouds based on their morphological type.
The presence of a significant number of head-tail clouds and their distribution 
in the region is discussed in the context of Magellanic System simulations. 
We suggest that ram-pressure stripping is a 
more important factor than tidal forces for the morphology and 
formation of the Magellanic Leading Arm and that different environmental conditions 
might explain the morphological difference between the Magellanic Leading Arm and 
Magellanic Stream. 
We also discuss a newly identified population of clouds that forms the LA IV 
and a new diffuse bridge-like feature connecting the LA II and III complexes.  
 
\end{abstract}

\section{INTRODUCTION\label{intro}}

Some atomic neutral hydrogen (\HI) concentrations surrounding our Galaxy have anomalous 
velocities that are forbidden by a simple Galactic rotation model.  
These so-called anomalous-velocity clouds contain \HI\ 
without a stellar counterpart. They can be classified into two velocity 
based groups: intermediate velocity clouds (IVCs; \citealp{MZ61,BT66}); 
high-velocity clouds (HVCs; \citealp{Muller63}).  
The classification of IVCs and HVCs is based on the deviation velocity, 
which is defined as the smallest difference between the velocity 
of the cloud and the Galactic rotational velocity  \citep{Wakker91}. 
High-velocity clouds are particularly interesting because 
they are thought to represent the flow of baryons in or out of the 
Galactic disk, which influences the formation and evolution of our Galaxy. 
Despite being important in the context of galaxy formation and evolution, 
their origin and physical characteristics are still under debate. 

Possible explanations of the origin of HVCs can be traced back 
to an early study by \citet{Oort66}. 
One of his hypotheses suggested that the HVCs have an extragalactic origin. 
This hypothesis received more recent support from 
\citet{Blitz99} with the argument that HVCs are dark matter dominated 
clouds in the Local Group with distances of hundreds of kiloparsecs. 
A similar study by \citet{BB99} also claimed that 
compact and isolated HVCs lie at extragalactic distances. 
Another popular HVC origin hypothesis is the Galactic fountain model, 
in which the gas is blown out of the disk by supernovae, cools and
then rains back down (see \eg, \citealp{Houck90}).
In this scenario, the fountain gas can rise as high as $\sim$10~kpc above the disk 
\citep{DA00}. 

The previous large-scale surveys, the Leiden--Argentine--Bonn 
Galactic \HI\ survey (LAB; \citealp{Kalberla05}) and the \HI\ Parkes All-Sky Survey 
(HIPASS; \citealp{Barnes01}) have provided opportunities to study 
HVCs on a global scale to assist in the understanding of their origin and physical 
properties (see \eg, \citealp{WW91,Putman02}). 
LAB covered the entire sky with an angular resolution 
of 36$\arcmin$ and a spectral resolution of 1.3~\kms. 
HIPASS was conducted with a better angular resolution (16$\arcmin$) 
but lower spectral resolution (18~\kms). 
A comprehensive catalog of Southern HVCs based on the HIPASS data 
was presented in \citet{Putman02}, hereafter P02. 
The catalog covers the high-velocity \HI\ sky south of declination +2\degr\ and 
within the Local Standard of Rest velocity ($V_{LSR}$) range of +500 to $-$500~\kms. 
It provides the spatial and kinematic 
distributions as well as the properties of high-velocity clouds. 
Even though the P02 catalog includes a complete census of HVCs 
south of declination +2$\degr$, 
the nature of the in-scan bandpass calibration technique filtered 
out some of large-scale structure of the Milky Way and the Magellanic System. 

Among HVC complexes, the Magellanic System is the most interesting given that 
it is the only closest extragalactic gaseous stream to our Galaxy. 
The Magellanic System consists of a coherent gas stream originating from the Magellanic Clouds (MCs), 
\ie, Magellanic Stream (MS) and Leading Arm (LA) \citep{Mathewson74}. 
The MS is trailing the MCs and has a complex filamentary structure. 
On the other hand, the LA is clumpy and dominated by three distinctive large complexes, 
namely the LA I, LA II and LA III \citep{Putman98,Bruns05}. 
An extended feature of the MS has recently been discovered by \citet{Nidever10}, 
which reveals the total length of the MS as $\sim$200\degr\ across the sky. 
Another recent report of several filaments  
that are aligned with the MS also suggests that MS is wider than 
previously thought \citep{Westmeier11}. 
   
The formation of the MS and LA is generally believed to have been caused by the tidal interaction 
between the Milky Way and Magellanic Clouds. 
Theoretical models with tidal stripping, gravitational and hydrodynamical interactions 
can reproduce global observed \HI\ column density and velocity distributions 
(see \eg, \citealp{Connors06,Mastropietro05}). 
However, these models do not provide a satisfactory explanation for 
the formation mechanism of the MS and LA. 
With the recent Hubble Space Telescope proper motion measurements of 
the MCs \citep{K06b,K06a}, 
a new unbound orbit for the MCs with a first passage scenario was proposed by \citet{Besla07}. 
The result is surprising given that the new orbit does not provide 
sufficient time for tidal and ram-pressure 
stripping mechanisms to produce the MS \citep{Stanimirovic08}. 
To circumvent the problem raised by the first passage scenario, \citet{Nidever08} 
proposed a new blowout hypothesis. They suggested that the supergiant shells in the 
dense southeast \HI\ overdensity region are blown out from the 
Large Magellanic Cloud (LMC) to larger radii 
where ram-pressure and/or tidal forces can be more easier to 
strip the gas and form the MS and LA. 
Nevertheless, 
with recent higher measurements of the Milky way's circular velocity 
(\eg, 251~\kms; \citealp{Reid04}) compared to the IAU standard of 220~\kms, 
there remains the distinct possibility that a multi-orbit scenario is plausible
\citep{SL09}.
Recent multi-orbit simulations have also reproduced the 
observed structure of the MS, including bifurcation 
of the two filaments, whilst remaining consistent with the proper motion data \citep{DB11,DB12}. 
However, none of the theoretical models to date have been able to accurately reproduce the
observed structure of the LA.  

To study the Magellanic System in detail, 
\citet{Bruns05} carried out a narrow-band Parkes \HI\ survey. 
In contrast to HIPASS, which was not designed to accurately measure 
low-velocity Galactic \HI\ gas, the survey was designed exclusively 
to study the Magellanic System.   
The Br\"{u}ns' survey has a similar angular resolution and spectral resolution 
to the Galactic All-Sky Survey (GASS; \citealp{MG09}; \citealp{Kalberla10}) (see \S2), 
but with limited sky coverage. Another high-resolution \HI\ study that 
concentrated on the northern tip of the MS was carried out by \citet{Stanimirovic08}. This 
study was part of the Galactic studies with Arecibo $L$-band Feed Array (GALFA).

The current work utilizes the GASS data for studying 
the general distribution and morphological types of HVC 
in the region of the Magellanic Leading Arm. The GASS data have better 
sky coverage than the Br\"{u}ns' survey and higher 
spectral resolution than HIPASS. The study of HVCs 
in the vicinity of the LA gives us clues to understand: (1) the 
formation of the LA; (2) the physical properties of the HVCs; and most importantly,  
(3) the role of infalling gas in the context of galaxy 
evolution and formation.       
In \S2, we describe the GASS data and the procedures  
for cloud search algorithms. We present the catalog and the general distribution 
of clouds in \S3 and \S4. Classification of the clouds and interpretation of 
the distribution for each group are given in \S5. 
Finally, we report on the new extended features of the Magellanic Leading Arm 
and discuss the implications of HVCs for the formation and origin of the LA in 
\S6. Conclusions are drawn in \S7.

\section{DATA\label{data}}

The neutral hydrogen data employed here are from GASS. 
This survey covers the entire Southern sky to declination $+1\degr$ and 
$V_{LSR}$ from $-$400 to +500~\kms.  
The data from the GASS second data release\footnote{http://www.astro.uni-bonn.de/hisurvey/gass/}
have been corrected for stray radiation, have 
an angular resolution of $\sim16\arcmin$, a brightness temperature ($T_{\rm B}$) sensitivity of 57~mK, 
a channel width of 0.82~\kms\ and a spectral resolution of 1~\kms. 
For a typical HVC of 15~\kms\ line width in our sample, the 1$\sigma$ \HI\ column density ($N_{\rm HI}$) 
sensitivity is $3.5\times10^{17}$~cm$^{-2}$.
We refer the reader to \citet{MG09} and \citet{Kalberla10} for a detailed description of 
the observing technique and data reduction. 

\begin{figure}
\plotone{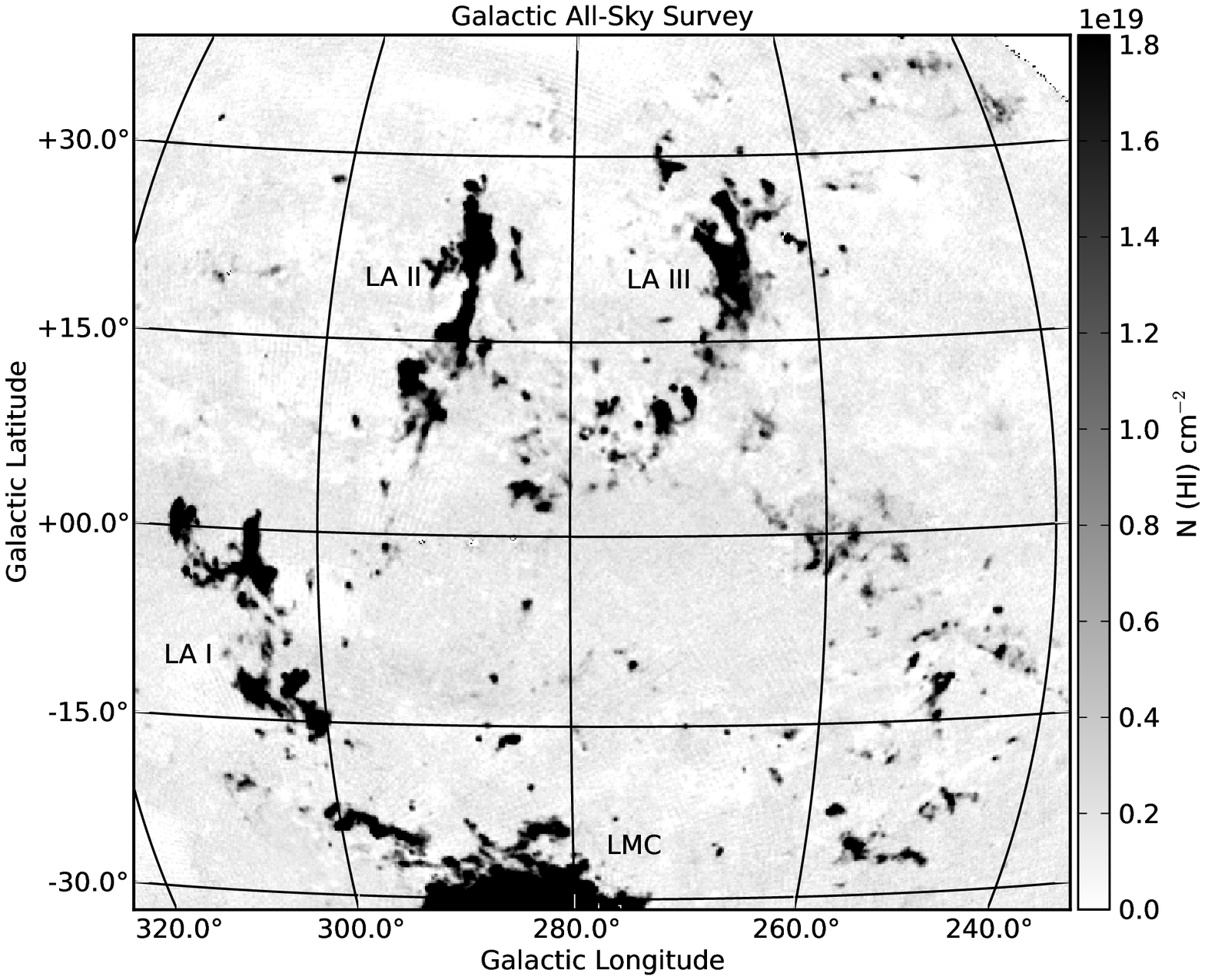}
\caption{The integrated \HI\ column density map of GASS in the 
region of the Magellanic Leading Arm. The \HI\ column density scale is 
0 to 1.8$\times10^{19}$~cm$^{-2}$. Locations of the Leading Arm complexes 
I, II, III and the Large Magellanic Cloud are labeled. \label{LA_mom0}}
\end{figure}

To create a catalog of HVCs in the region of the Magellanic Leading Arm, we 
extracted a GASS data cube within the area of $-30\degr \lesssim$~$b$~$\lesssim +40$ and 
$240\degr \lesssim$~$l$~$\lesssim 315\degr$, and covered 
the velocity range of 0 to 450~\kms. 
Before performing any cloud search algorithm, we examined the data cube and 
determined the velocity range that solely contain Galactic 
\HI\ emission, 0~\kms~$\leq$~$V_{LSR}$~$<$~150~\kms. 
Beyond this velocity range, there is a mix of Galactic \HI\ emission and 
possible \HI\ emission that is associated with the MCs. 
\citet{Bruns05} analyzed the HVCs in the region of the LA 
by using different velocity ranges within certain Galactic latitudes. 
Due to the difficulty in distinguishing the Galactic \HI\ and \HI\ emission 
originating from the MCs, 
we deliberately masked out the regions of $-20\degr \lesssim$~$b$~$\lesssim +20\degr$ 
and $l < 310\degr$ between 150 and 190~\kms\ for our analysis. 
Other small regions of Galactic \HI\ emission in higher velocity
channels, as determined by eye,  were also masked.
An integrated \HI\ column density map of HVCs in the region of the LA 
over the velocity interval of 150 to 450~\kms\ is shown in Figure~\ref{LA_mom0}.

\subsection{Source Finding \label{duchamp}}

We employed the source finding software, 
{\it Duchamp}\footnote{Available at http://www.atnf.csiro.au/computing/software/duchamp/} V1.1.13, 
developed by \citet{Whiting12}.
It is a 3-dimensional source finding software that provides flexibility 
for the user to control all relevant input parameters. 
To enhance the detectability of fainter sources, it implements 
optional noise reduction routines, such as the 
$\grave{a}$ {\it trous} wavelet reconstruction technique \citep{SM02} and 
the spatial or spectral smoothing routine. 
A full description of {\it Duchamp} is given in \citet{Whiting12}. 
Here we describe the basic flow of the program 
with the applied input parameters: 
\begin{itemize}
\item Spectral channels with Milky Way emission, 
0~\kms~$\leq$~$V_{LSR}$~$<$~150~\kms, were flagged (see \S 2) and 
were excluded when performing the search.

\item The cube was reconstructed via the $\grave{a}$ {\it trous} wavelet 
reconstruction method. It determined the amount of structures at 
various scales, and random noise was removed from 
the cube based on a user defined threshold.  

\item We did not use the spectral or spatial smoothing  
to remove the random noise because tests had shown that the 
$\grave{a}$ {\it trous} wavelet reconstruction method yielded 
a better source detection rate. 

\item A fixed threshold of $\sim2\sigma$ above the background 
noise (57~mK) was specified for the source finding. 
We did not adopt the auto-threshold determining scheme of {\it Duchamp}.

\item {\it Duchamp} searched for sources one channel at a time 
using the defined threshold. Sources were confirmed only if 
they extend to a minimum of 5 channels in velocity space and 10 pixels spatially.

\item Subsequently, 
detections were compared to earlier detected sources and 
either combined with a neighboring source or 
added to the list.  
\end{itemize}

We also performed the source finding using {\it Clumpfind}. 
{\it Clumpfind} is a cloud finding algorithm developed to quantify the fragmentation or 
clumpiness of molecular clouds \citep{WGB94}. 
The routine contours the data with user defined root mean square noise of the 
observations and interval, then searches for peaks of emission to 
locate the clumps at each contour level and splits any blended clumps with 
a friends--of--friends algorithm.
We chose to adopt {\it Duchamp} for this study as it breaks sources
into fewer components. 

\section{CATALOG\label{cat}}

\begin{figure}
\epsscale{0.9}
\plotone{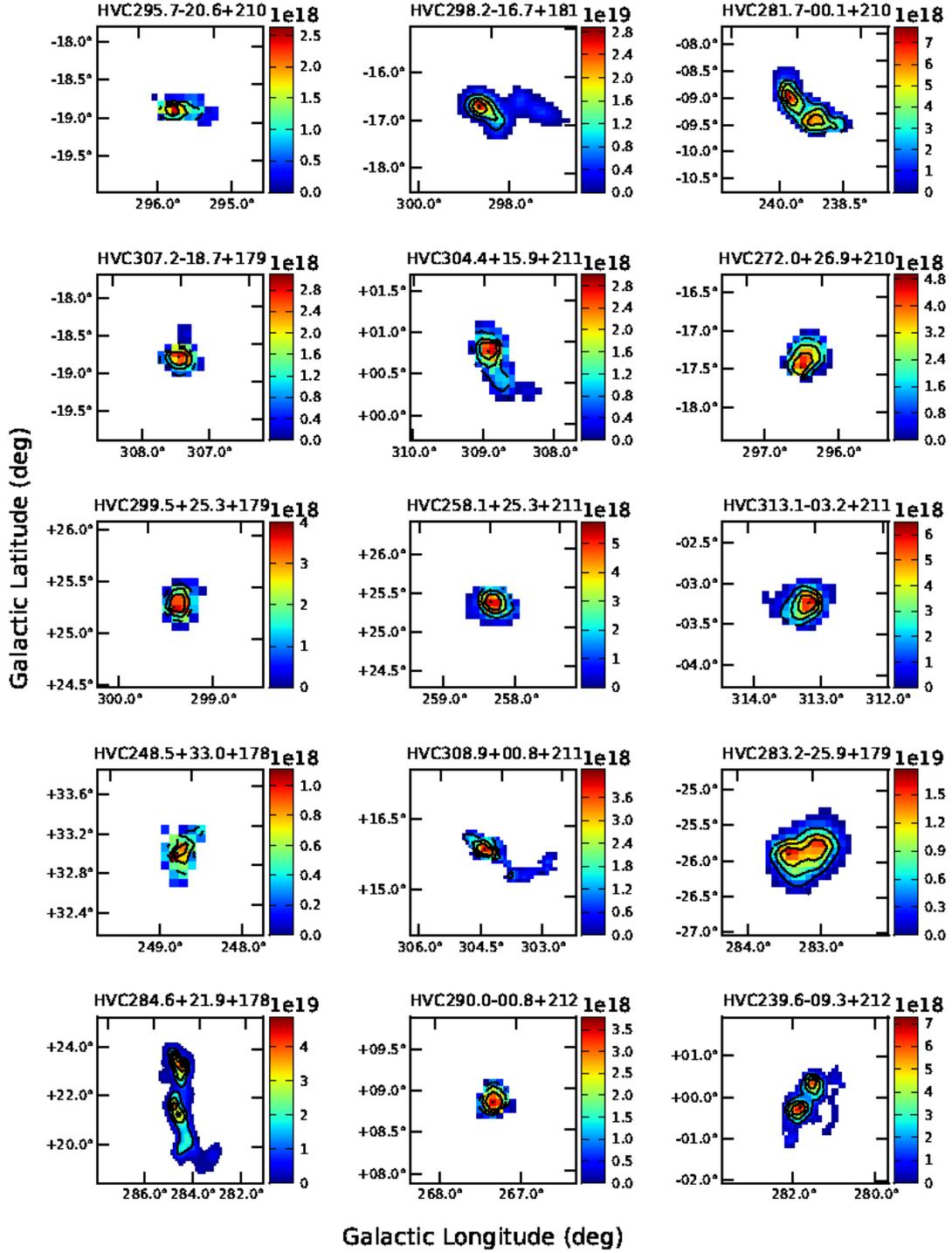}
\caption{Examples of integrated \HI\ column density maps of individual sources. 
Contours and colors representing the 
\HI\ column density scale are shown. \label{subcubes_mom0}}
\end{figure}

\begin{figure}
\epsscale{0.9}
\plotone{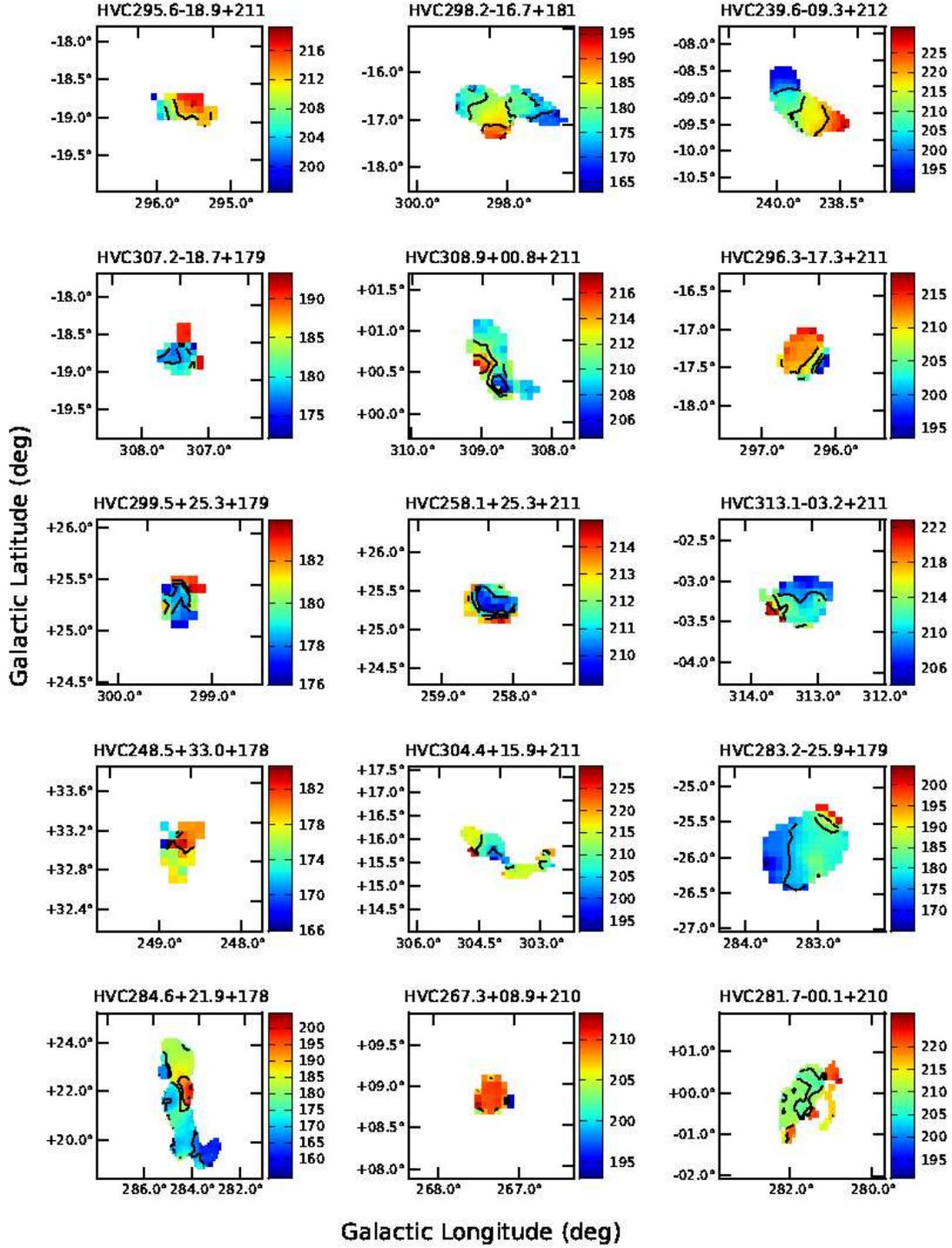}
\caption{Examples of velocity field maps of individual sources. 
Contours and colors representing the LSR velocity are shown. 
\label{subcubes_mom1}}
\end{figure}

We present the basic information of the sources in Table~\ref{catalog}. 
The catalog includes the object identification number in column 1; 
the designation with a prefix of HVC for high-velocity clouds and GLX for galaxies 
followed by the Galactic longitude, Galactic latitude and 
the velocity in the Local Standard of Rest reference frame ($V_{LSR}$) of the source in column 2; 
$V_{LSR}$ in column 3; 
the velocity in the Galactic Standard of Rest reference frame ($V_{GSR}$), 
defined by $V_{GSR}=220$~cos~$b$~sin~$l$~$+V_{LSR}$; 
the velocity in the Local Group Standard of Rest reference frame  ($V_{LGSR}$), 
defined by $V_{LGSR}=V_{GSR}-62$~cos~$l$~cos~$b$~$+40$~sin~$l$~cos~$b$~$-35$~sin~$b$ \citep{BB99} in columns 5--6; 
velocity FWHM, measured at 50$\%$ of peak flux in column 6; 
integrated flux, peak $T_{\rm B}$ and peak $N_{\rm HI}$ in columns 7--9;
semi-major axis, semi-minor axis and position angle in columns 10--12;  
warning flag in column 13; and classification in column 14 (see \S5). 
Most of the parameters are derived from {\it Duchamp}, with the exception of 
angular sizes and peak \HI\ column density, which are determined independently. 
Sources that straddle our velocity boundary and masked Milky Way emission boundary 
are included in the catalog but their physical parameters cannot be
determined accurately and are therefore not listed in the table. We present the 
examples of integrated \HI\ column density maps and LSR velocity maps of 
individual cloud in Figures~\ref{subcubes_mom0} and \ref{subcubes_mom1}.   

With the procedures described in \S\ref{duchamp}, 
we found 838 sources with {\it Duchamp}. 
False detections caused by background artifacts were eliminated from the initial search.
The positions of final detected sources were subsequently examined using 
the NASA/IPAC Extragalactic Database with a $16\arcmin$ search radius to identify any galaxy. 
The final count includes a total of 419 HVCs and 
12 galaxies in the region of the Magellanic Leading Arm. 
We determined the peak $N_{\rm HI}$ by locating the brightest pixel in the integrated 
\HI\ column density map of each source. 
To determine the angular size of the HVCs, we used 2-dimensional 
Gaussian and elliptical fitting, which gave the semi-major axis, semi-minor axis 
and position angle. 
A detailed investigation and discussion of advantages and disadvantages for 
both fitting methods, applied to molecular cloud catalogs, is given in \citet{Kerton03}. 
We elect to use the results 
from the 2-dimensional Gaussian fitting for the catalog. 
Due to the inaccuracy of the position angle in some cases, 
caution is needed in interpreting this parameter.   

\begin{figure}
\begin{center}
\includegraphics[angle=-90,scale=0.5]{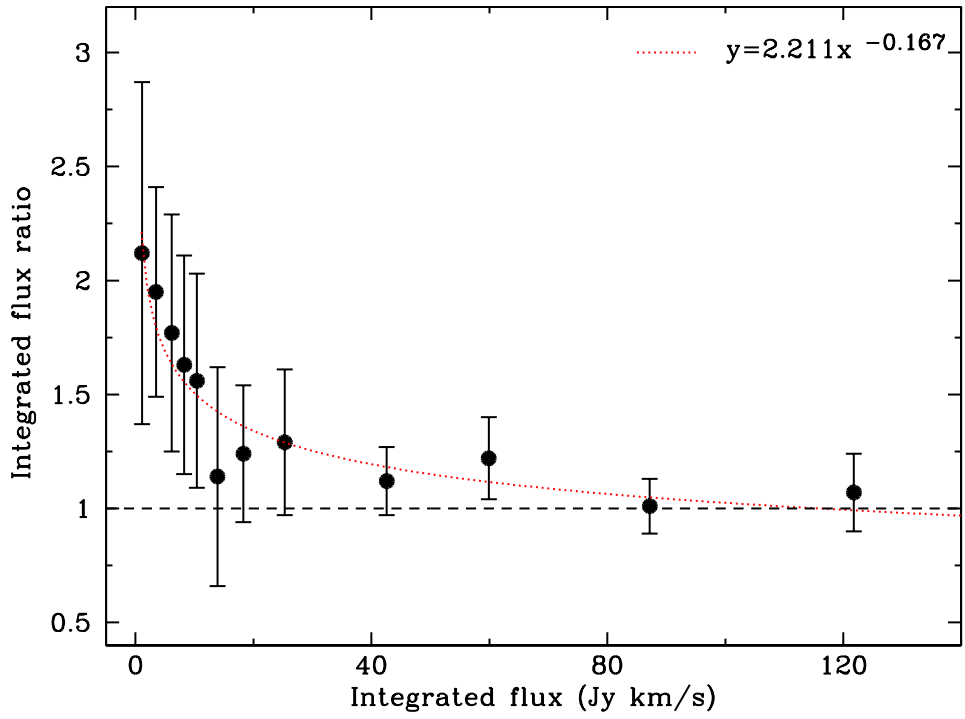}
\caption{Ratio of ``true'' integrated flux ($F_{\rm int}$) to the integrated flux ($F^{'}_{\rm int}$) 
measured by $Duchamp$ as a function $F^{'}_{\rm int}$. 
The red dotted line is the fit to the data points, which represents our estimated correction 
to the $Duchamp$ values. 
\label{iflux_comp}}
\end{center}
\end{figure}

The reliability of {\it Duchamp} for parametrization of sources has been extensively tested 
on artificial sources of various parameters by \citet{Westmeier11}. 
Their tests show that the integrated flux ($F_{\rm int}$) measured by {\it Duchamp} is 
systematically too low for faint sources. 
To demonstrate how this systematic error affects the integrated flux of real sources 
found in our catalog, we measured the $F_{\rm int}$ with a better
parametrization algorithm. This stand-alone parametrization algorithm uses 
the position output from {\it Duchamp} to find sources in the data cube. 
An integrated map is created and an ellipse fitted to the source. 
Subsequently, the ellipse is grown in size until the measured 
integrated flux reaches a maximum. The $F_{\rm int}$ is then measured within the 
final ellipse.
In Figure~\ref{iflux_comp}, we show that the ratios of our measured $F_{\rm int}$ to 
$F^{'}_{\rm int}$ as measured by {\it Duchamp} as a function of measured {\it Duchamp} $F^{'}_{\rm int}$ 
in various bins. 
Parametrization of large and extended or close kind confused sources is 
challenging (T. Westmeier, priv communication). 
Thus, these sources are excluded from the comparison. 
The red, dotted line is a fit to the data points, which represents the underestimated  
factor for a given $F^{'}_{\rm int}$ as measured by {\it Duchamp}. 
We conclude that {\it Duchamp} produces accurate measurements of $F^{'}_{\rm int}$ for sources 
with $F^{'}_{\rm int} \gtrsim 80$~Jy~\kms\ in our catalog. 
We correct all the $F^{'}_{\rm int}$ values measured by {\it Duchamp} based on the fitting function.  

Corrections are not necessary for peak $T_{\rm B}$ (fitted by {\it Duchamp}) and
peak $N_{\rm HI}$ (measured from the column density map). 
However, small systematic errors in an order of 5--10$\%$ of derived values may also 
be present in these parameters. 
The {\it Duchamp} determined velocity FWHM values, however, 
are generally very accurate (see Figure~8 of \citealp{Westmeier11}).
We will compare our catalog with the catalog compiled from a recent study 
by \citet{Venzmer12}, hereafter V12, in Appendix.

\subsection{Comparison with P02 Catalog \label{compcatalogs}}

Here we compare our catalog to the P02 HVC catalog.  The P02 catalog
is based on HIPASS data, which have been reprocessed with the MINMED5
method to recover extended emission \citep{Putman00}.  The catalog
covers the entire sky in the declination range 
$-90\degr$ to $+2\degr$, and the velocity range of $+90$~\kms\
$\lesssim |V_{LSR}| < +500$~\kms.  The $|V_{LSR}|$$< 90$~\kms\ limit
does not exclude all emission associated with the Milky Way at low
Galactic latitudes, and an additional constraint of deviation velocity
was applied to their selection criterion.  The search was performed
via an automated friends-of-friends HVC finding algorithm of
\citet{HBB02}.

\begin{figure}
\plotone{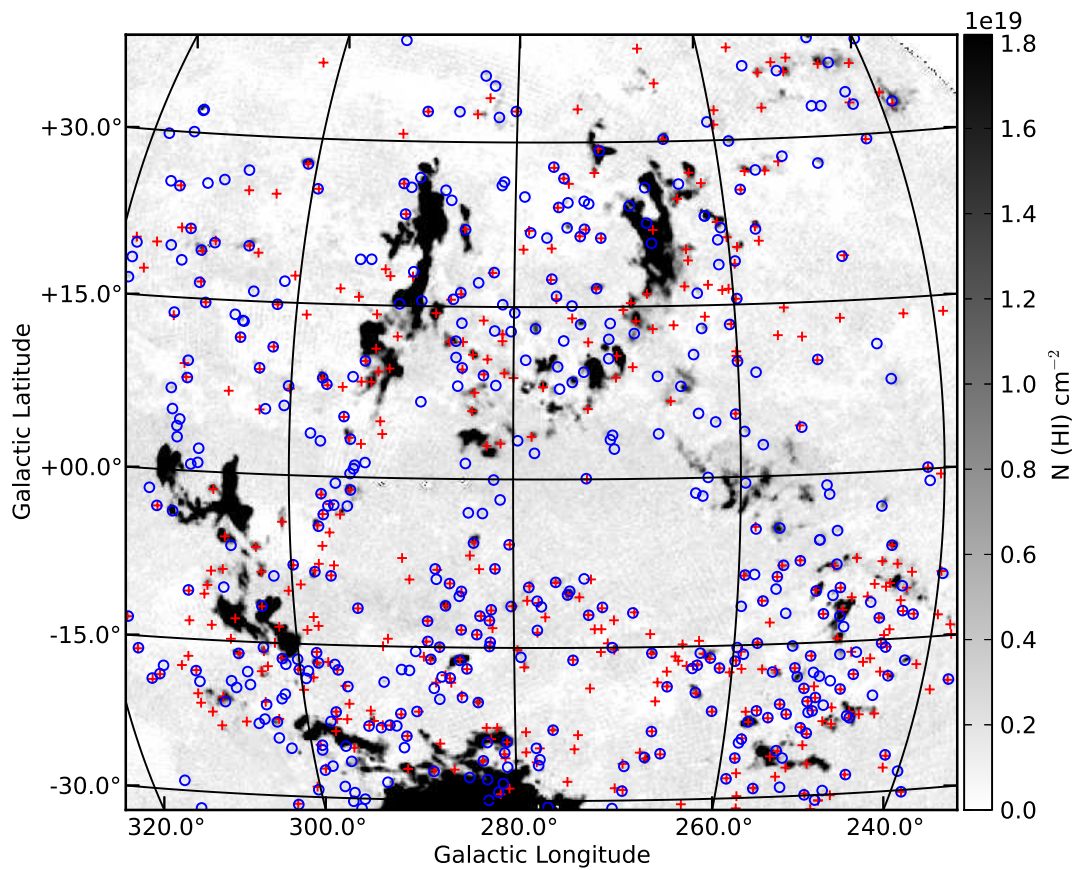}
\caption{On-sky distribution of the 431 sources detected by {\it Duchamp} (blue circles) 
and the 448 sources detected by \citet{Putman02} (red pluses) in the region of the Leading Arm. 
The integrated \HI\ column density map of Figure~\ref{LA_mom0} is shown. The 
sources from \citet{Putman02} are within the same velocity range as the catalog presented in this paper. 
Comparison between the two catalogs is discussed in \S\ref{compcatalogs}. 
\label{duchampvshipass}}
\end{figure}

\begin{figure}
\plotone{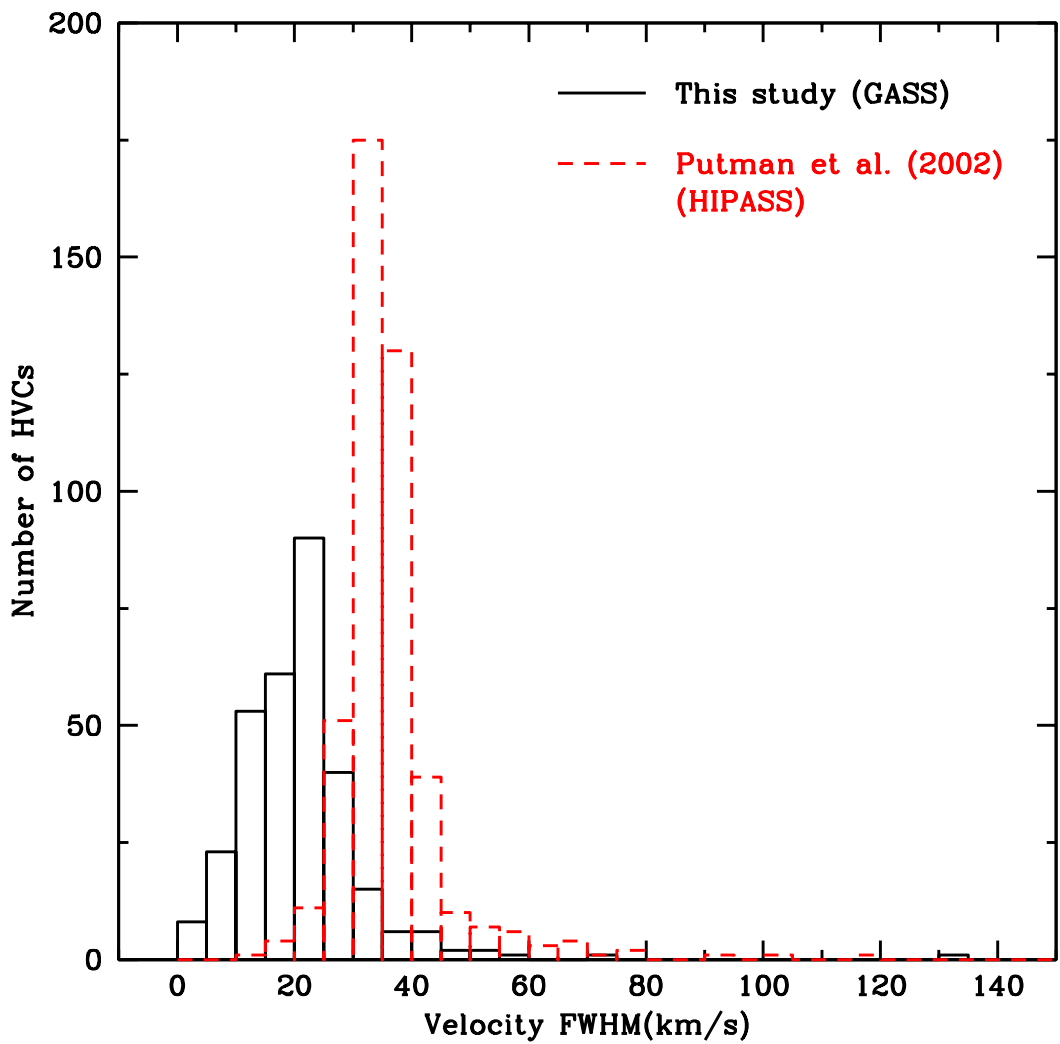}
\caption{Histograms of velocity FWHM of HVCs in P02 (dashed line) and this study (solid line). 
Excluded from the plot are: 30 clouds that extend below $V_{LSR} = 150$~\kms; 
71 clouds that extend into the masked Milky Way 
emission boundary; 10 clouds that extend into the spatial edge of the image and galaxies. 
A larger number of narrow-line width HVCs are recovered in this study as compared to P02. 
The difference is mainly due to the higher spectral resolution of GASS.
\label{wvel_dist}}
\end{figure}

In Figure~\ref{duchampvshipass}, we show a comparison between HVCs
from the P02 catalog that fall within our searched GASS survey volume
and in our catalog.  The total number of identified HVCs in this
region is 419 for our catalog and 448 for the P02 catalog.  We point
out that our catalog includes 30 clouds which straddle our velocity
boundary at $V_{LSR}$= 150~\kms\ and 71 clouds which extend over to
the boundary of masked Milky Way emission region.  About 230 HVCs are
identified to be the same cloud in the two catalogs.  The differences
between the catalogs are due to: (a) degrees
of breaking up or merging clouds in extended complexes, 
for which {\it Duchamp} tends to merge clouds while 
the friends-of-friends algorithm employed by P02 tends to break up clouds; (b) the
superior brightness sensitivity of HIPASS compared to GASS (9~mK
per 15 \kms\ compared with 15~mK per 15\kms\ in GASS), resulting in
more faint sources to be detected in P02; (c) the excellent spectral
resolution of GASS has allowed us to resolve narrow-line HVCs in the
region, for which the spectral lines would have been smeared out in
the case of coarser spectral resolution of HIPASS; (d) GASS has a
better baseline coverage at the Galactic plane than HIPASS, resulting
in more sources to be detected near the Galactic plane.  We present a
velocity FWHM distribution of both catalogs in Figure~\ref{wvel_dist}.
HVCs that straddle our masked Milky Way emission boundaries and
galaxies have been excluded in the plot.  This figure shows that the
majority of HVCs in the P02 catalog possesses larger velocity FWHM
(30--35~\kms) than ours (10--25~\kms).

\subsection{Completeness of the catalog \label{completeness}}

We adopted the same method as described in \citet{Begum10} 
to evaluate the completeness of our catalog. Ten fake clouds 
with various input parameters were injected into our data cube at random locations. 
The injected fake clouds were modeled with the Gaussian function. 
Input parameters of LSR velocity, peak $T_{\rm B}$, velocity FWHM, position angle, 
angular size in term of semi-major and semi-minor axes were randomly 
selected from various ranges. Ranges of input parameters 
were determined based on the cloud properties in our catalog. 

We performed four sets of simulations with fake clouds being injected 
into the data cube between 150~\kms\ and 450~\kms\ LSR velocities:   

\begin{itemize}

\item Bright clouds with a narrow velocity line width: $T_{\rm B}$ = 1.1--3.0~K, 
velocity FWHM = 5--15~\kms, semi-major axis = 0.3$\degr$--0.7$\degr$, 
semi-minor axis = 0.2$\degr$--0.4$\degr$.

\item Bright clouds with a broad velocity line width: $T_{\rm B}$ = 1.1--3.0~K, 
velocity FWHM = 16--30~\kms, semi-major axis = 0.5$\degr$--1.0$\degr$, 
semi-minor axis = 0.2$\degr$--0.4$\degr$.

\item Faint clouds with a narrow velocity line width: $T_{\rm B}$ = 0.14--1.0~K, 
velocity FWHM = 5--15~\kms, semi-major axis = 0.3$\degr$--0.7$\degr$, 
semi-minor axis = 0.2$\degr$--0.4$\degr$.

\item Faint clouds with a broad velocity line width: $T_{\rm B}$ = 0.14--1.0~K, 
velocity FWHM = 16--30~\kms, semi-major axis = 0.3$\degr$--0.7$\degr$, 
semi-minor axis = 0.2$\degr$--0.4$\degr$.  

\end{itemize}

Injected bright and faint clouds with a narrow velocity 
line width were fully recovered. The detection rates are 
8/10 and 9/10 for faint and bright clouds with a broad velocity line width, respectively. 
The bright cloud with a broad velocity line width was missed because it 
was merged with another neighboring cloud. This is not uncommon considered that 
a cloud with a broader velocity line width has a higher chance in overlapping 
with another cloud in a crowded region, and hence merged into a larger cloud complex. 
Faint clouds with a broad velocity line width can be lost in the 
background noise. 
To summarize, it is likely that our catalog missed faint, broad velocity line width clouds 
but $Duchamp$ is relatively reliable in detecting narrow line width clouds for the studied region.

\section{General Distribution and Properties\label{distribution}}

\begin{figure}
\plotone{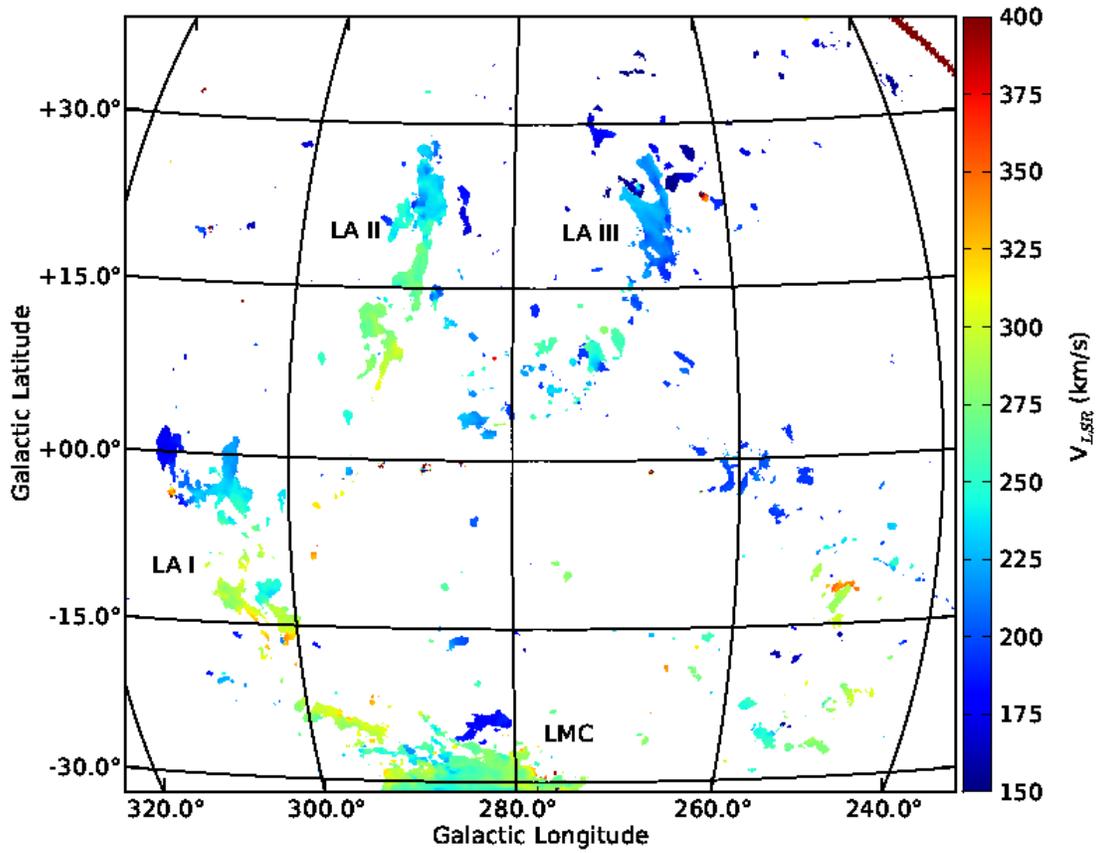}
\caption{The velocity field map (first moment map) in the LSR velocity reference frame. 
The color bar represents the velocity range from 150 to 400~\kms. 
The Leading Arm complexes and Large Magellanic Cloud are labeled. \label{mom1_vlsr}}
\end{figure}

\begin{figure}
\plotone{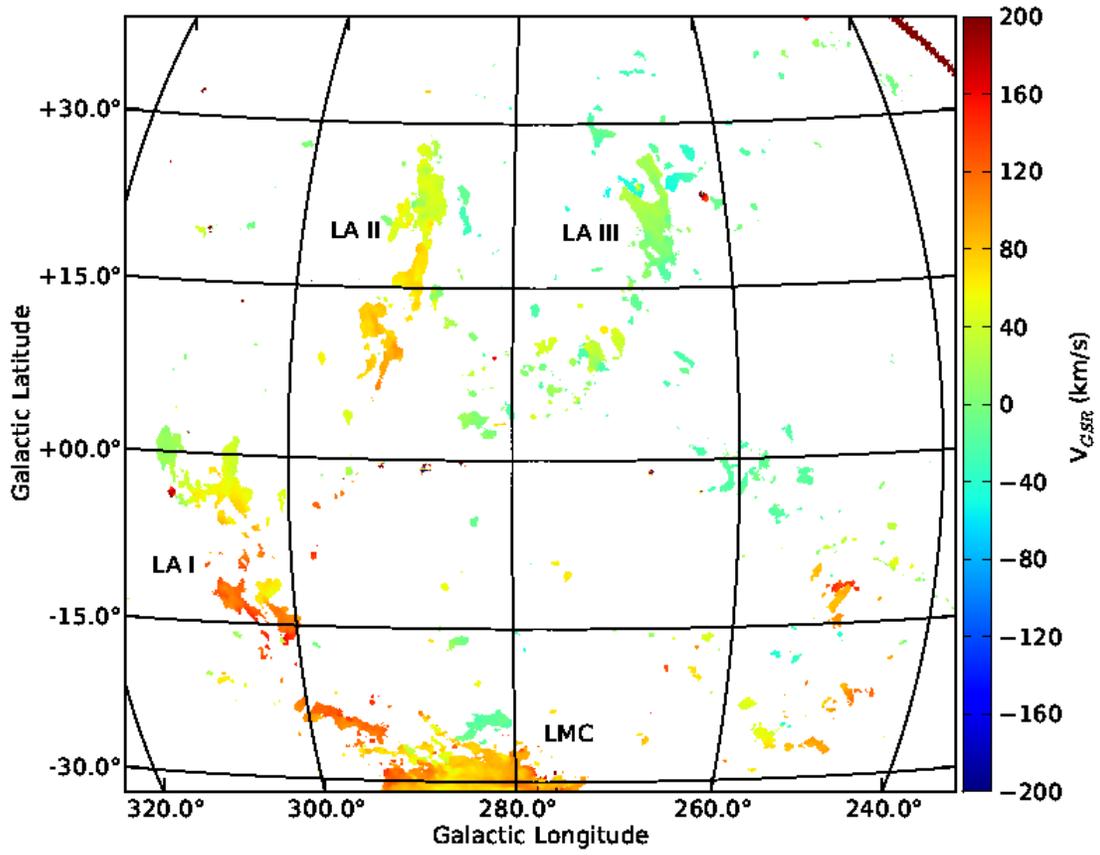}
\caption{Same as Figure~\ref{mom1_vlsr}, except in the GSR velocity reference frame 
with a velocity range of $-$200 to +200~\kms.\label{mom1_vgsr}}
\end{figure}

The integrated \HI\ column density map of HVCs in the region of the 
LA shows a significant concentration of HVCs 
between Galactic longitude 240\degr\ to 
260\degr\ and Galactic latitude $-$30\degr\ to +0\degr\ 
(see Figure~\ref{LA_mom0} and \S6). 
This population appears to be clumpy but has a few larger, more 
complex clouds ($\sim3\degr$--$5\degr$ in angular size). Velocity field maps 
in the LSR and GSR reference frames, 
as shown in Figures~\ref{mom1_vlsr} and \ref{mom1_vgsr}, indicate large velocity 
gradients for the LA I and LA II complexes.   
Faint or thin filamentary structures of complex clouds 
are not visible in these maps but can be seen in the moment maps of individual sources 
(\eg, see Figures~\ref{subcubes_mom0} and \ref{subcubes_mom1}). 
We assume all of these HVCs are originated from the MCs and 
are associated with the LA due to their close proximity on the sky to the 
MCs and their similar range of velocities. 
The following statistical analysis excludes galaxies and objects otherwise flagged 
in the catalog.  

\begin{figure}
\plotone{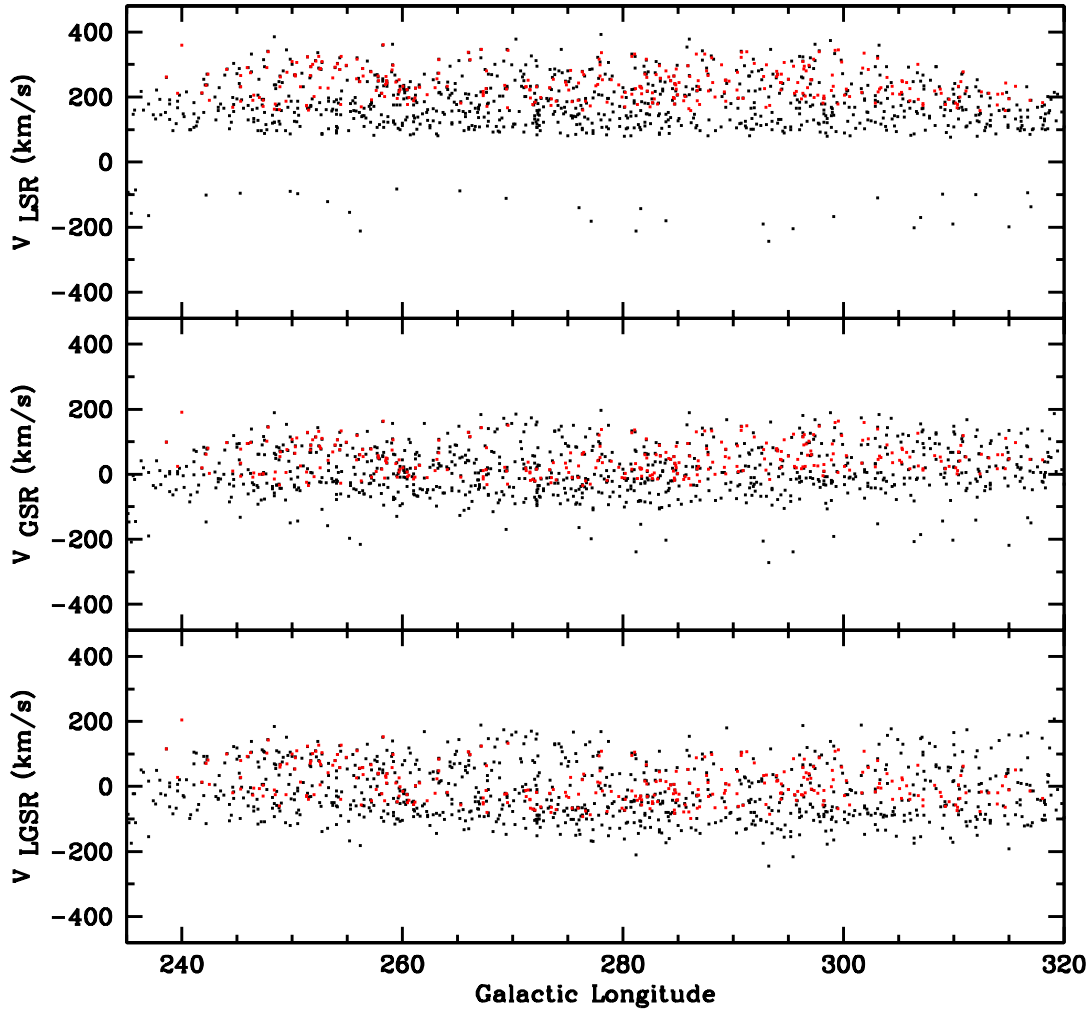}
\caption{Kinematic distributions of HVCs in the $V_{LSR}$, $V_{GSR}$ and $V_{LGSR}$ reference 
frames versus Galactic longitude (from top to bottom). 
The black and red dots represent HVCs in P02 
and our catalog, respectively. 
Excluded from the plot are: 30 clouds that extend below $V_{LSR} = 150$~\kms; 
71 clouds that extend into the masked Milky Way 
emission boundary; 10 clouds that extend into the spatial edge of the image and galaxies. 
\label{vel_l}}
\end{figure}

\begin{figure}
\plotone{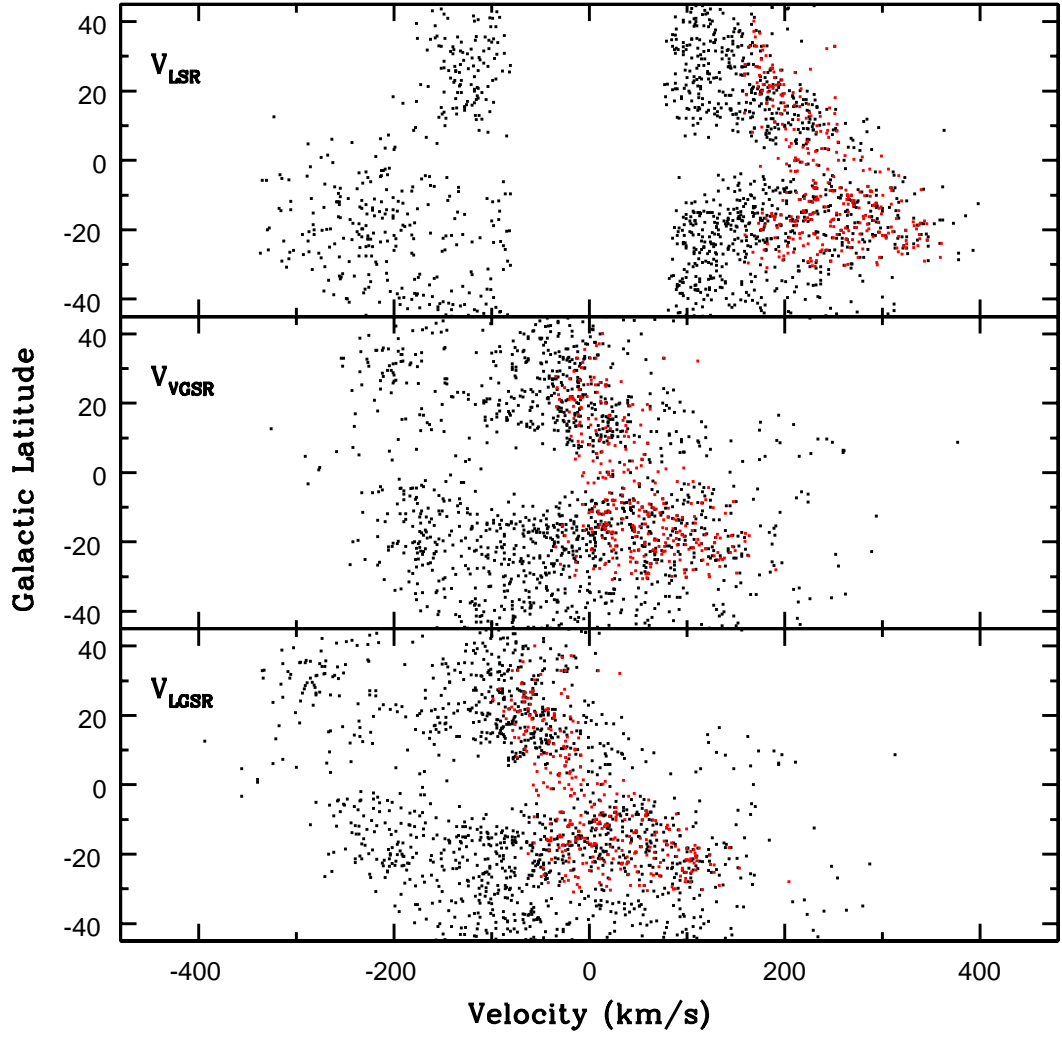}
\caption{Same as Figure~\ref{vel_l}, except showing velocities versus Galactic latitude.\label{vel_b}}
\end{figure}

We present the kinematic distributions of HVCs in $V_{LSR}$, $V_{GSR}$ and $V_{LGSR}$ 
versus Galactic longitude and latitude in Figures~\ref{vel_l} and \ref{vel_b}, 
respectively.  A comparison between HVCs in the P02 catalog (black dots) 
and ours (red dots) is also shown. The main difference is that 
our catalog only includes sources with $V_{LSR} > 150$~\kms. 
In the top panel of Figure~\ref{vel_l} there are more 
positive than negative LSR velocity HVCs in the given Galactic longitude range. 
The overall velocity distribution of all P02 HVCs in Galactic and Local Group reference 
frames (middle and bottom panels of Figure~\ref{vel_l}) has
nearly equal number of HVCs with positive and 
negative velocities. The HVCs in our catalog are evenly distributed across 
the Galactic longitude between 240\degr\ and 320\degr\ in all reference frames. 

The kinematic distributions of our HVCs in Galactic latitude is slightly 
different than for the HVCs in the P02 catalog. The top panel of Figure~\ref{vel_b} 
shows the lack of identified clouds in the range of 150~\kms $< V_{LSR} <$ 190~\kms\ and 
$-15\degr < b < +15\degr$ in our catalog. The lack of clouds in this region is 
caused by the way we constructed the data cube by deliberately masking out 
most of the emission in this velocity range to avoid any contamination from the 
Galactic \HI\ emission (see \S2). 
There is a lack of clouds at higher Galactic latitude as the velocity increases 
in all reference frames. 

\begin{figure}
\plotone{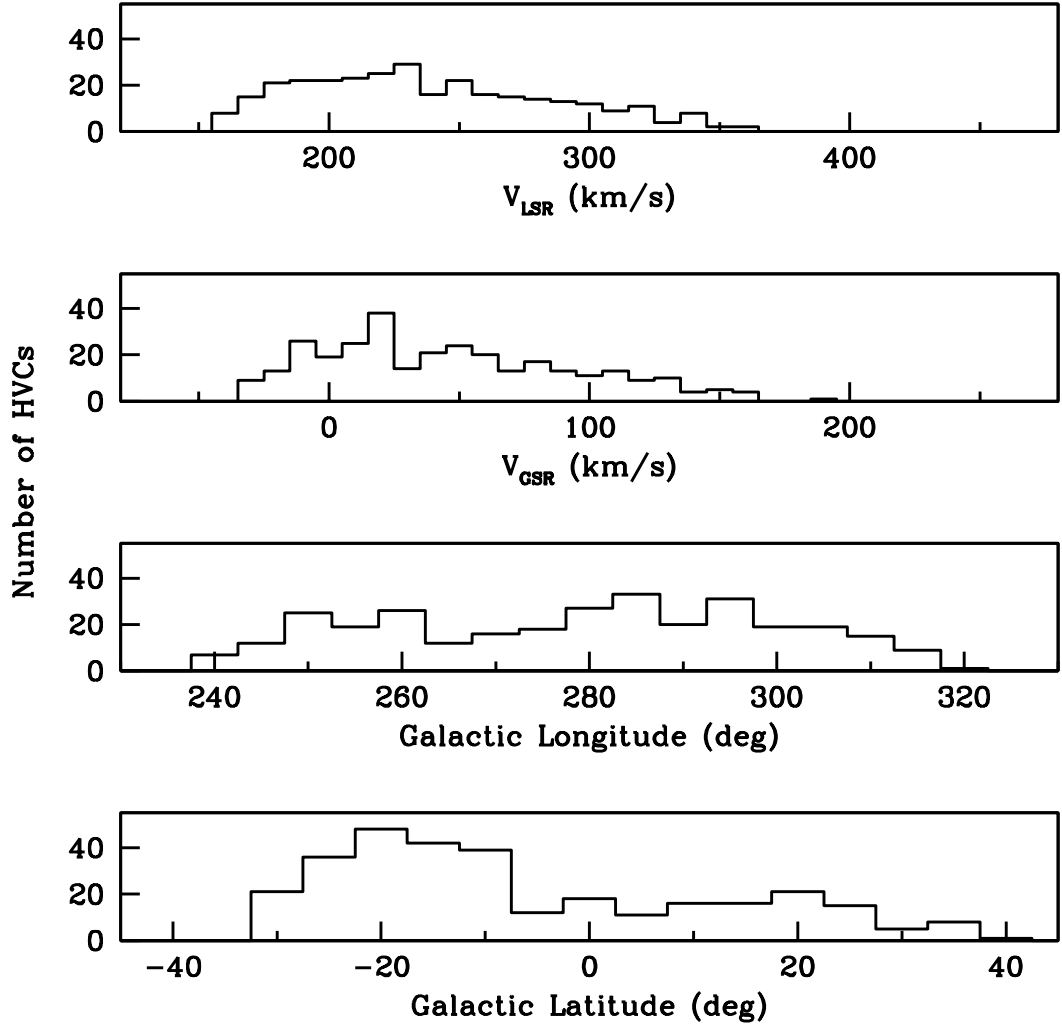}
\caption{Histograms of $V_{LSR}$, $V_{GSR}$, Galactic longitude and Galactic latitude of HVCs identified 
in the GASS data,from top to bottom, respectively. 
Excluded from the plot are: 30 clouds that extend below $V_{LSR} = 150$~\kms; 
71 clouds that extend into the masked Milky Way 
emission boundary; 10 clouds that extend into the spatial edge of the image and galaxies. 
\label{vlb_hist}}
\end{figure}

In Figure~\ref{vlb_hist}, we show histograms of $V_{LSR}$, $V_{GSR}$, 
Galactic longitude and latitude (from top to bottom panels). 
The median $V_{LSR}$ and $V_{GSR}$ for the HVCs in the catalog are 232~\kms\ and 42~\kms, 
respectively, which is consistent with the mean radial velocity of the LA \citep{Bruns05}. 
We find that the number of clouds declines gradually 
as $V_{LSR}$ increases above 200~\kms. 
Nearly the same number of clouds per 5\degr\ bin is found between 
Galactic longitude 250\degr\ and 300\degr (third panel). 
The total number of clouds below the Galactic plane outnumbers those above the Galactic plane, 
with a large fraction agglomerated between Galactic latitude $-25$\degr\ and $-10$\degr\ 
(bottom panel), where the region is closer to the LMC. 

\begin{figure}
\plotone{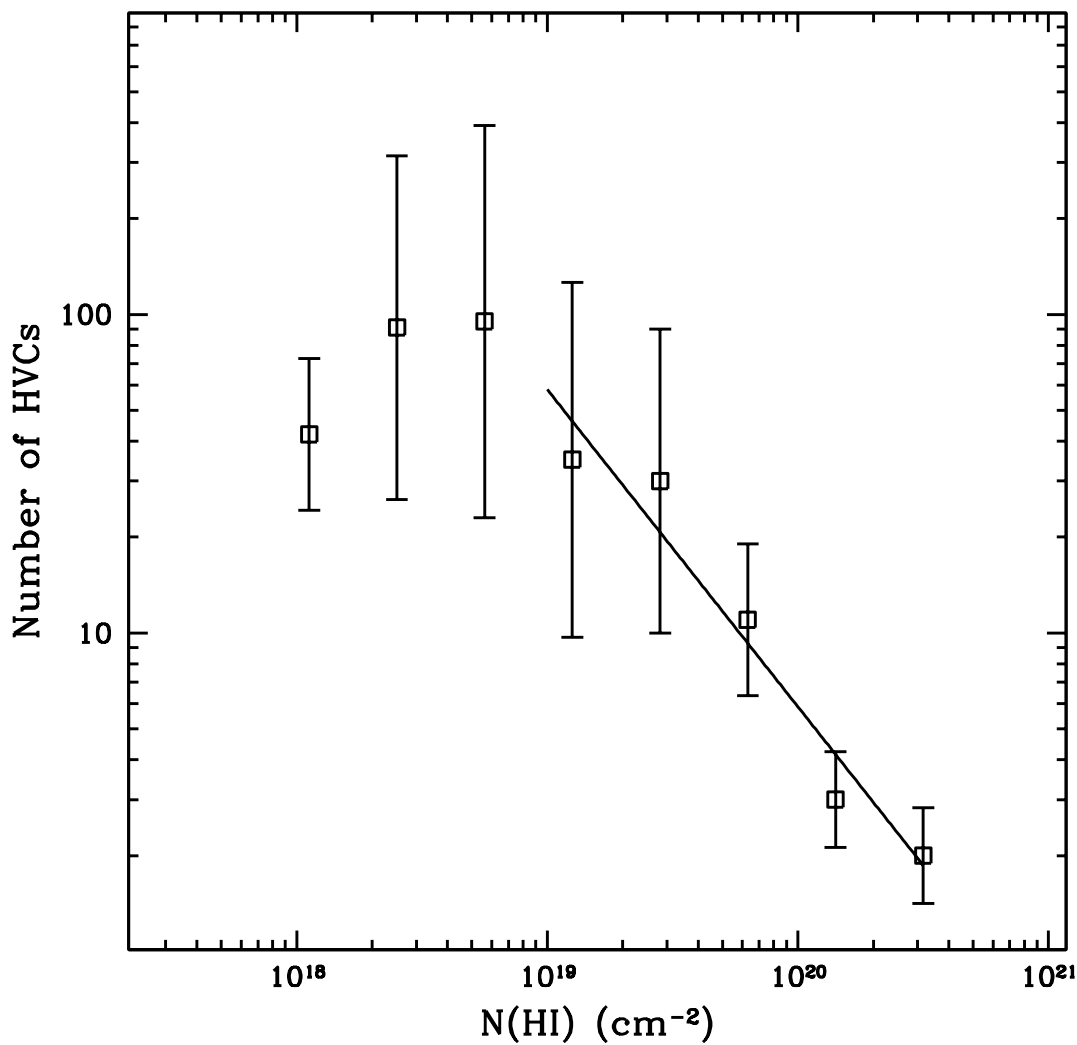}
\caption{The peak \HI\ column density distribution of HVCs in our catalog. 
A slope of $-$1.0 is derived from least square fitting of the data points above $10^{19}$~cm$^{-2}$
in the log-log plane, which corresponds to 
distribution function, $f$($N_{\rm HI}$)$\propto N_{\rm HI}^{-2.0}$.
Excluded from the plot are: 30 clouds that extend below $V_{LSR} = 150$~\kms; 
71 clouds that extend into the masked Milky Way 
emission boundary; 10 clouds that extend into the spatial edge of the image and galaxies. 
\label{distfunc}}
\end{figure}

In Figure~\ref{distfunc}, we show the distribution function of peak
\HI\ column density, which can be described by a power law:
$f$($N_{\rm HI}$)$\propto N_{\rm HI}^\alpha$.  The distribution shows
that high column density clouds are rare.  The turn-over
at the low column density end of the distribution indicates that the
population is limited by the survey sensitivity
(1$\sigma=3.5\times10^{17}$~cm$^{-2}$). A linear function was fitted
in the log--log space and a negative slope of $-1.0$ was determined.
This yields the final form of the distribution function as $f$($N_{\rm HI}$)$\propto N_{\rm HI}^{-2.0}$. 
Comparing the distribution function of peak $N_{\rm HI}$ in the LA region with 
the MS (see Figure~10 of \citealp{Putman03}), we find that both distributions 
turn-over simultaneously at the low column density end but the slope 
is steeper for their distribution function ($\alpha=-2.8$). 
This implies that clouds of all \HI\ masses in the MS contribute 
significantly to the total \HI\ mass of the Magellanic System.

\section{High-Velocity Cloud Morphological Classification\label{classification}}

\begin{figure}
\plotone{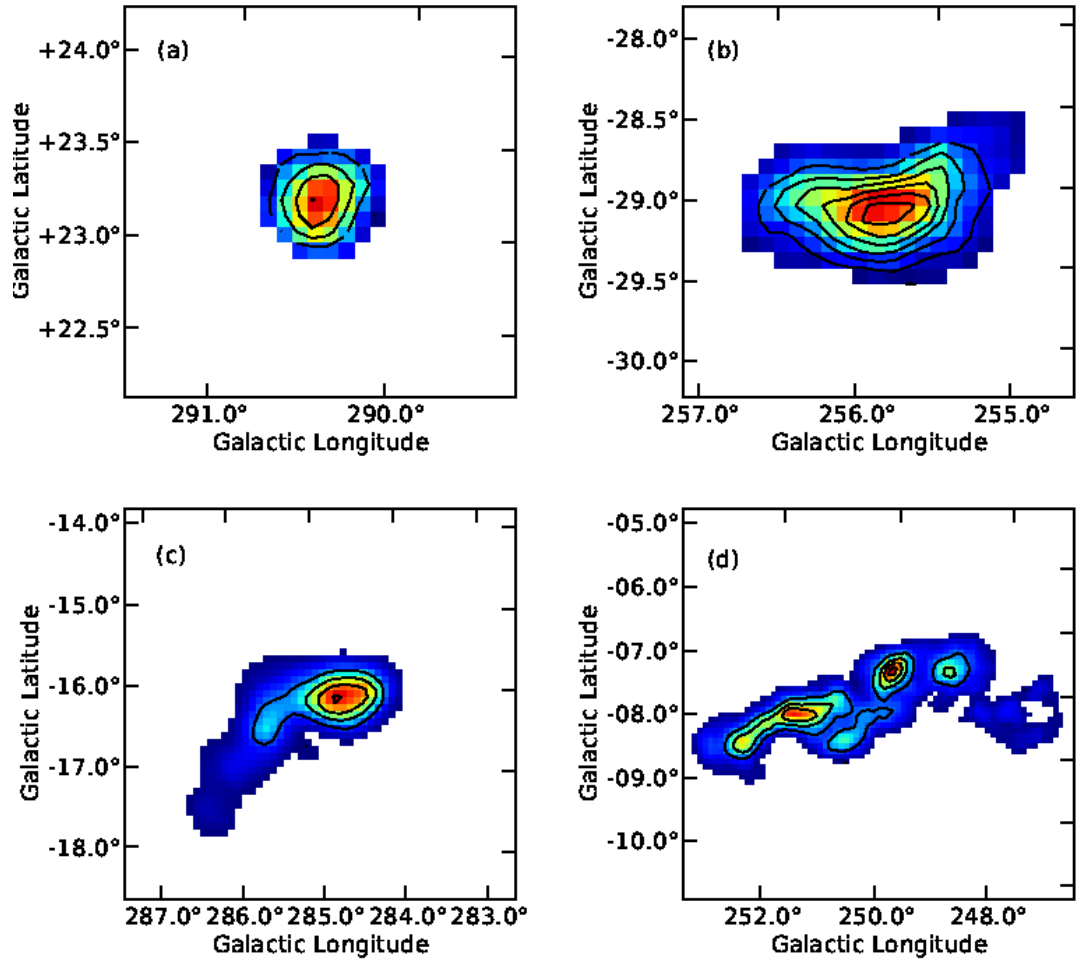}
\caption{Examples of cloud morphological types: (a) symmetric cloud; 
(b) bow-shock shaped cloud; (c) head-tail cloud; (d) complex and irregular cloud.
\label{diffshapes}}
\end{figure}

Different shapes of HVCs have been identified in the past, head-tail
clouds in particular have been studied extensively
(see \eg, \citealp{Bruns00}; \citealp{Westmeier05}; \citealp{Putman11}). 
Examining the integrated \HI\ column density and 
velocity field maps of each cloud, 
we can classify the HVCs into 5 groups: 
(1) clouds with head-tail structure and with velocity gradient (HT); 
(2) clouds with head-tail structure but without velocity gradient (:HT); 
(3) bow-shock shaped clouds (B);
(4) symmetric clouds (S); and 
(5) irregular/complex clouds (IC). 
In Figure~\ref{diffshapes}, we show different morphological type of clouds 
in our catalog. An analysis of groups 1--4 is given in the following subsections.

\subsection{Head-Tail Clouds\label{HTclouds}}

Traditionally, a head-tail cloud is defined as a cloud which appears to be 
cometary with a compressed head trailed by a relatively diffuse tail; 
and a clear column density gradient is visible \citep{Bruns00}. 
In this study, we find that some head-tail clouds consist of an additional 
clump of diffuse gas or have a kink in the tail, slightly more complex 
than the traditional head-tail clouds. 
This structure suggests a fraction of the gas is being ripped off from the main 
condensation when it interacts with the surrounding halo gas.

Among the head-tail clouds, some show a velocity gradient, 
which is generally also associated with a column density gradient. 
Such a velocity gradient is another possible indicator for the detection 
of distortion caused by the interaction between clouds and an ambient medium 
(\eg, see PSM11; \citealp{Bruns00}). 
Detection of a velocity gradient strongly depends on the spectral 
resolution of the data. 
Because of the high spectral resolution of GASS, in contrast to HIPASS, 
it is feasible to measure velocity gradient in addition to the 
\HI\ column density gradient when classifying head-tail clouds. 
We divide the head-tail clouds into two types (group 1 $\&$ 2) 
and analyze them separately. 

The total number of head-tail clouds is 100 ($\sim25\%$ of the sample), 
with typical head and tail column density 
different by a factor of 5 (typically $\Delta N_{\rm HI}\sim 4\times10^{18}$~cm$^{-2}$). 
60$\%$ (61/100) of the head-tail clouds show a clear velocity gradient.  
A wide range of velocity differences between the head and tail 
($\sim$5--25~\kms) is detected. 
We find that this particular group of head-tail clouds with velocity gradient 
consists of two subgroups, with the velocity of the head either leading (pHT) 
or lagging (nHT) the tail. 
The number in the subgroups is about the same, 30 and 31, respectively. 

\begin{figure}
\begin{center}
\includegraphics[scale=0.45]{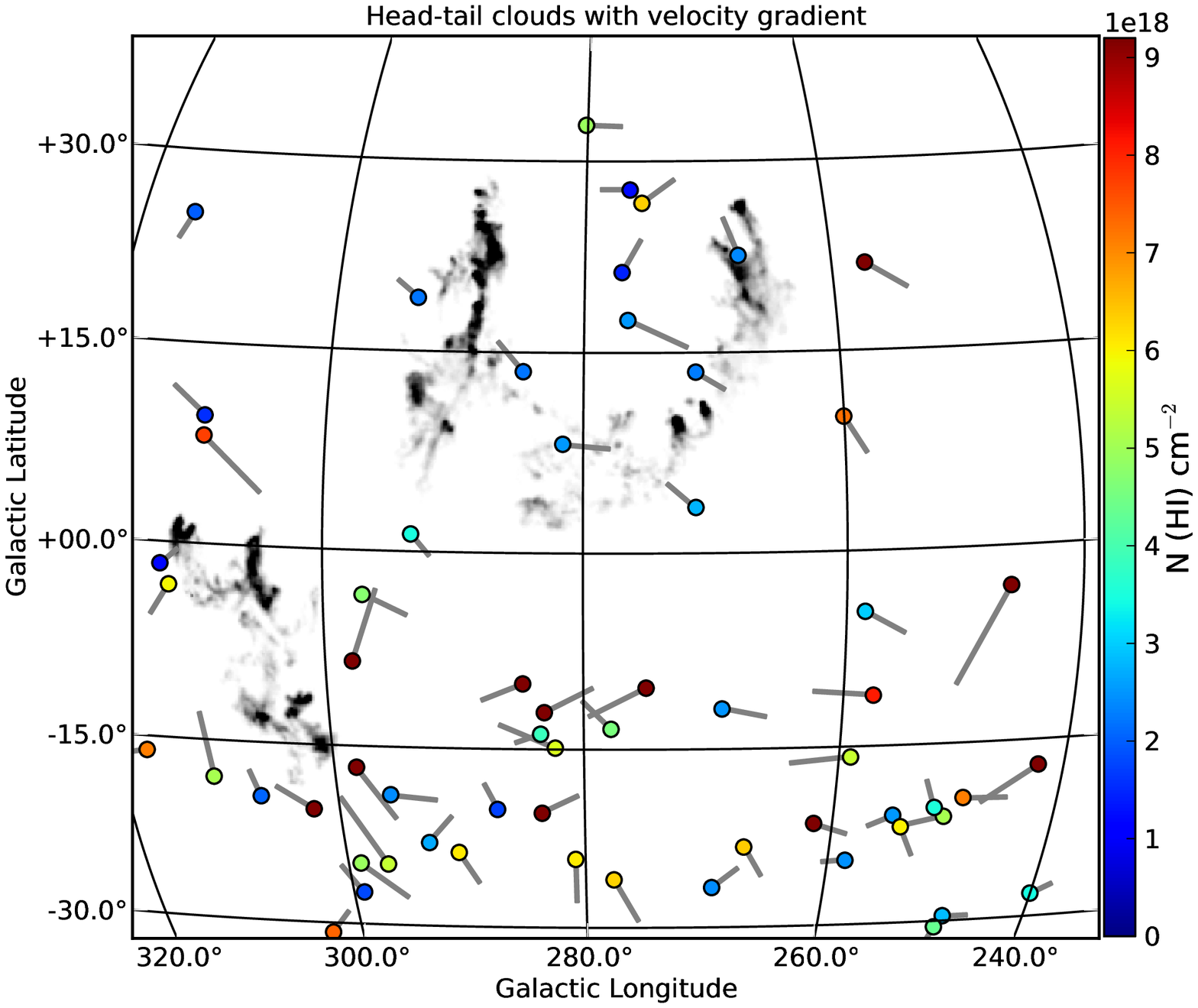}
\includegraphics[scale=0.45]{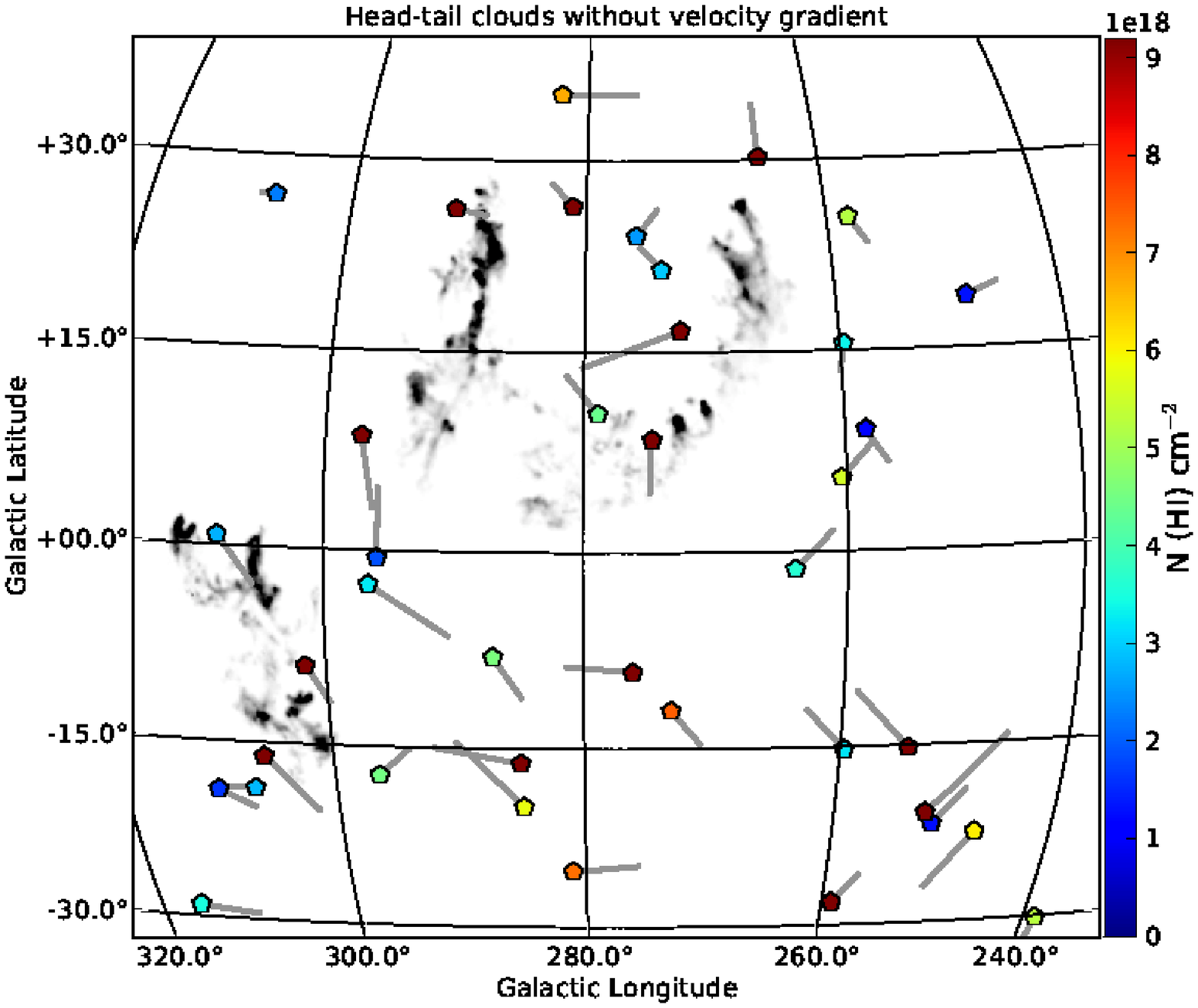}
\caption{Sky distribution of identified head-tail clouds with velocity gradient (top panel) and 
without velocity gradient (bottom panel) in the region of the Leading Arm. The colors indicate
the peak \HI\ column density of each head-tail cloud according to the color bar scale on the right side. 
The head and tail have been enlarged 
from its original size in these plots. The derived position angle of head-tail clouds has been 
adjusted manually to represent the actual pointing direction whenever necessary (see \S5.1). 
\label{HT_NHI}}
\end{center}
\end{figure}

In Figure~\ref{HT_NHI}, we show the peak \HI\ column density distributions 
of head-tail clouds with velocity gradient (top panel) and 
without velocity gradient (bottom panel) 
in the region of the LA. 
The pointing direction of the head-tail clouds is also presented, 
with the head and tail having been enlarged for better visibility. 
As discussed in \S3, the accuracy of the position angle is subject to 
the fitting methods and complexity of the cloud shape. 
To better represent the pointing direction of the clouds,  
we visually inspected each head-tail cloud and 
manually adjusted the Gaussian fit position angle whenever necessary. 

The head-tail clouds appear  
to be pointing in a random direction regardless whether they belong to group 1 or group 2. 
In contrast, the study by PSM11 found that 
the majority of head-tail clouds in the region of the LA 
point in the general direction of the North Galactic Pole, consistent 
with the general motion of the Magellanic System. 
The different conclusions may partly be due to the differences in selection criteria: 
(a) the velocity gradient is not well measured in PSM11; 
(b) only compact, isolated HVCs (CHVCs) are searched for head-tail structure in PSM11;
(c) the selection of head-tail clouds is a subjective process. 
We have about factor of two 
more head-tail clouds with velocity gradient than PSM11 in the same region. 
The implication of 
this random motion for the formation of the LA and 
its interaction with the Galactic halo will be discussed in \S6.2.    

\begin{figure}
\epsscale{0.7}
\plotone{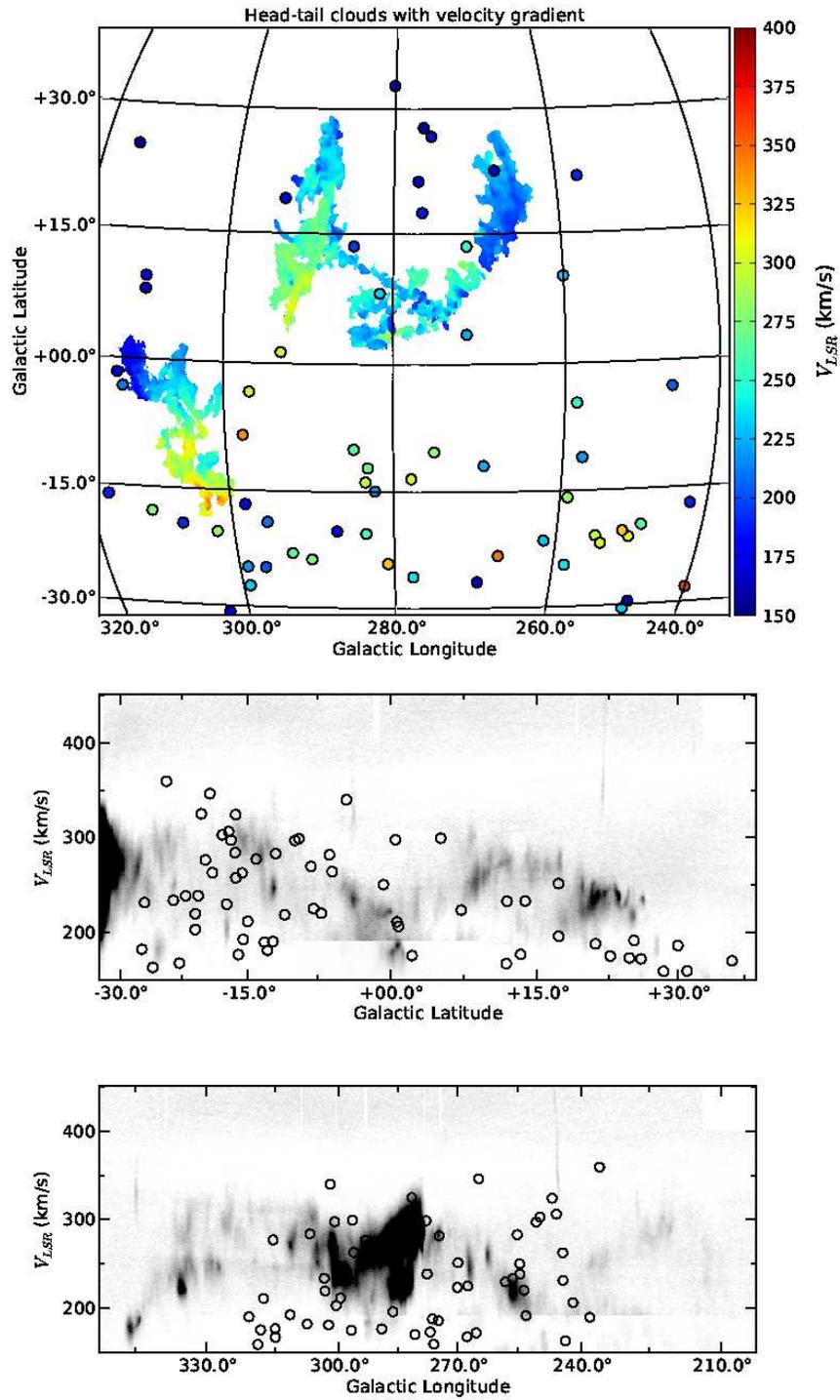}
\caption{The top panel shows the sky distribution of 
identified head-tail clouds with velocity gradient (HT) in the region 
of the Leading Arm. 
The colors represent the $V_{LSR}$ of each head-tail cloud according 
to the color bar scale on the right side. 
The bottom two panels show the HT clouds superimposed on position-velocity maps.
\label{HT_gradvel}}
\end{figure}

\begin{figure}
\epsscale{0.7}
\plotone{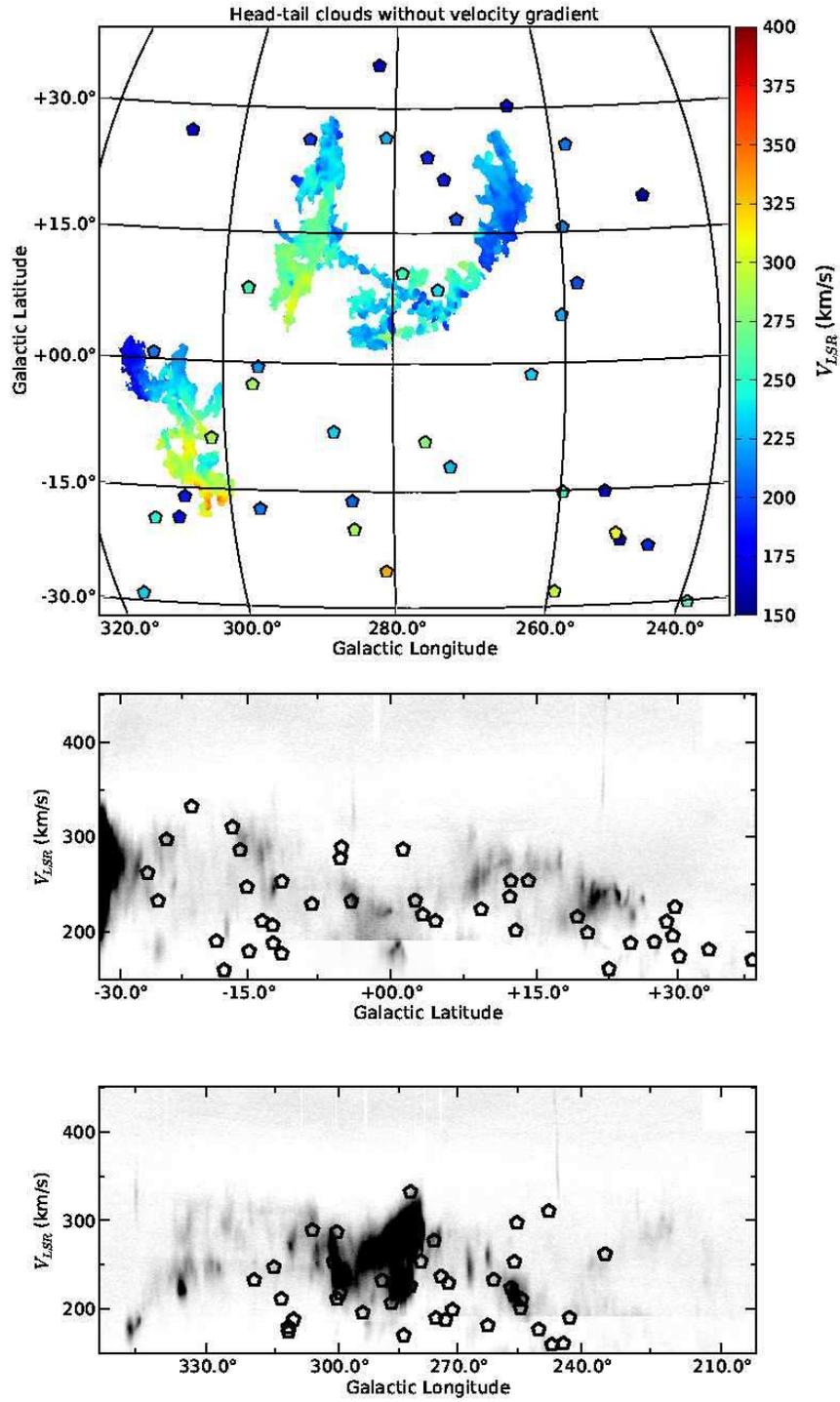}
\caption{Same as Figure~\ref{HT_gradvel}, except showing head-tail clouds without 
velocity gradients. \label{HT_nogradvel}}
\end{figure}

Examining the peak \HI\ column density distributions of both head-tail groups, 
we find that they  
populate the entire range of \HI\ column densities and are spread over the entire region. 
The distribution in $V_{LSR}$ for both groups is shown 
in Figures~\ref{HT_gradvel} and \ref{HT_nogradvel}, in which a dichotomy is found 
above and below the Galactic plane. 
Above the Galactic plane, majority of the HT and :HT clouds possess 
$V_{LSR} < 225$~\kms. Below the Galactic plane, it is populated by HT clouds with 
a wide range of $V_{LSR}$. 
   
\begin{figure}
\begin{center}
\includegraphics[scale=0.5]{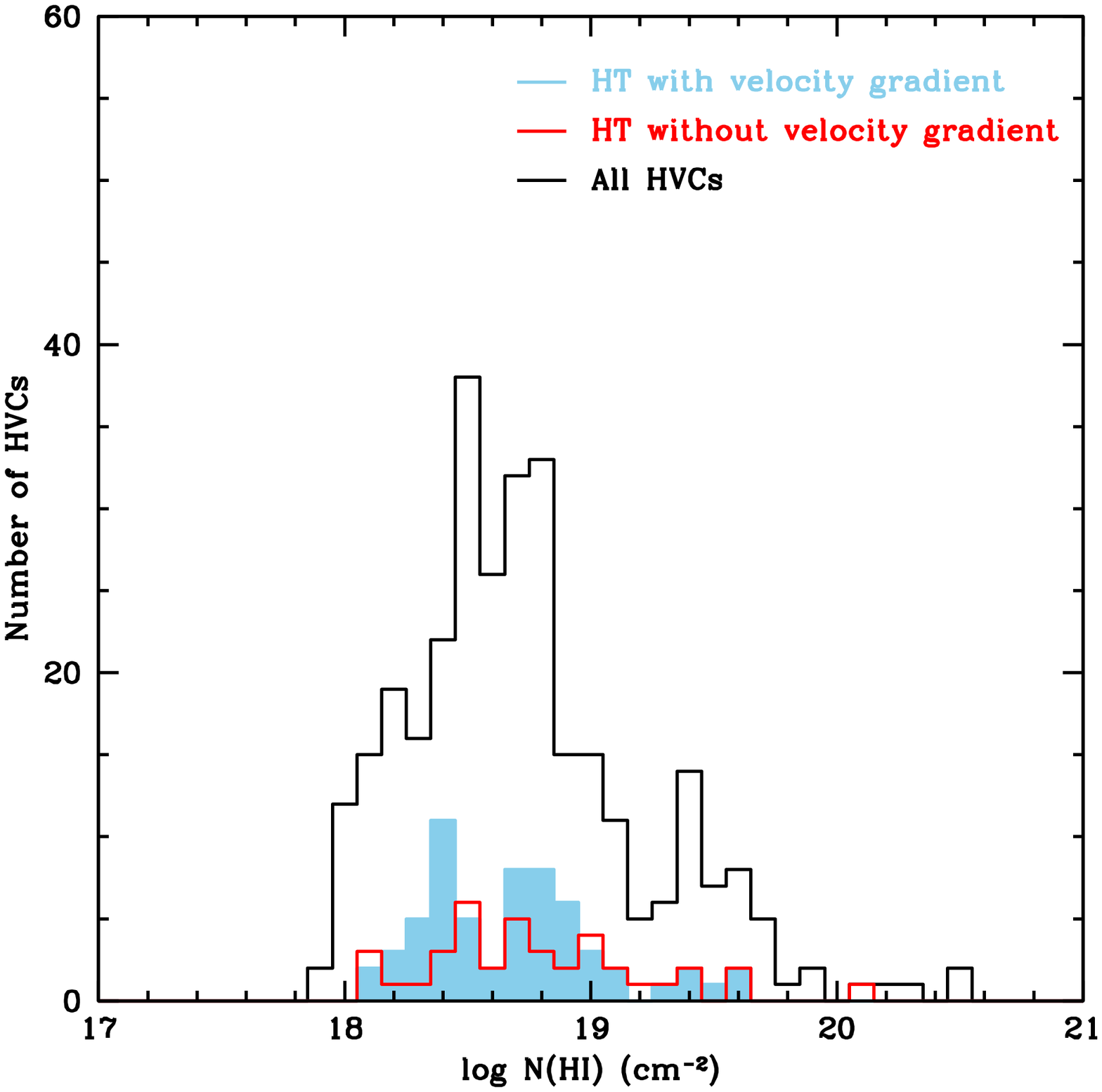}
\includegraphics[scale=0.5]{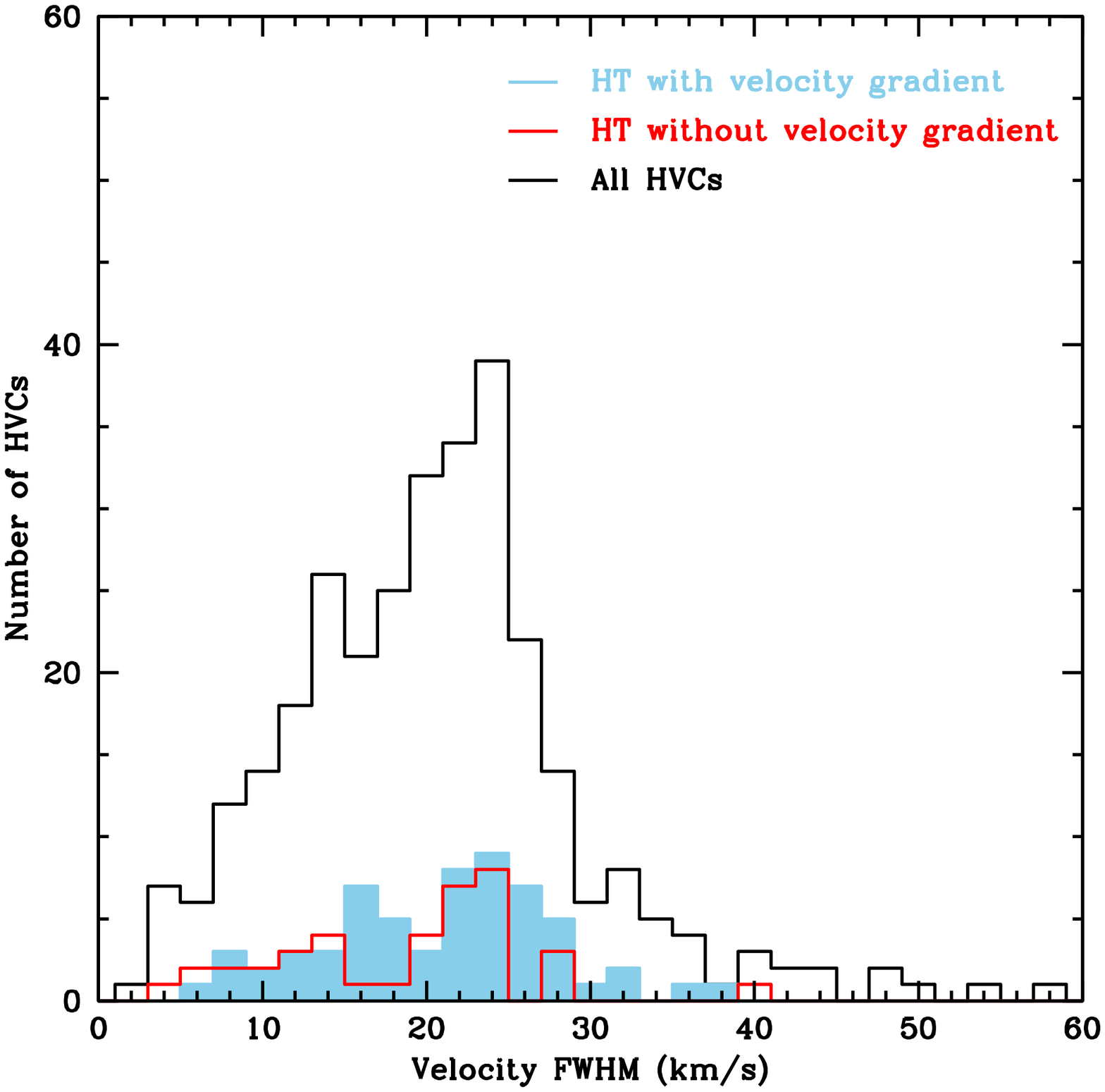}
\caption{Histograms of peak \HI\ column density (top panel) and velocity FWHM (bottom panel) 
of HVCs in this catalog. The black, blue and red represent all HVCs, head-tail clouds with velocity 
gradient and head-tail clouds without velocity gradient identified in GASS data, respectively. 
Only clouds with velocity FWHM less than 60~\kms\ are being plotted in the histogram. 
\label{hist_HTcolvel}}
\end{center}
\end{figure}

The distributions of peak \HI\ column density and velocity FWHM of 
these two head-tail groups as compared to the 
general population of HVCs in this study are presented in 
Figure~\ref{hist_HTcolvel}.  
The black, blue and red histograms represent 
all HVCs, head-tail clouds with velocity gradient (HT) and head-tail clouds
without velocity gradient (:HT) clouds, 
respectively. The peak \HI\ column density distributions are fairly similar in the cases of 
all HVCs and HT clouds,  
with the majority in the range of 18.4 $< \log$ (N$_{\rm HI}$/cm$^{-2}) <$ 18.6 (top panel). 
This result is different from the recent study by PSM11, 
where the majority of head-tail clouds 
possess $\log N_{\rm HI} > 19.0$. This is a factor of 2.5 higher 
in column density than those found in our sample. 
This is caused by the difference in 
the overall peak \HI\ column density distribution and 
number of detected narrow line width clouds. 
The larger number of narrow velocity line width clouds 
in this study than P02 is not due to the false detections (see \S 3.2). 
$Duchamp$ tends to merge clouds rather than breaking them into smaller clumps, 
which would result in a larger velocity line width for a merged cloud than smaller clumps.
The peak \HI\ column density distribution for :HT clouds is rather
flat, most likely because of the small sample size.
The velocity FWHM distributions are asymmetric with a peak at 22~\kms\ in all cases (bottom panel). 
Both velocity FWHM distributions of HT and :HT clouds extend out to $\sim40$~\kms. 
 
\begin{figure}
\begin{center}
\includegraphics[scale=0.43]{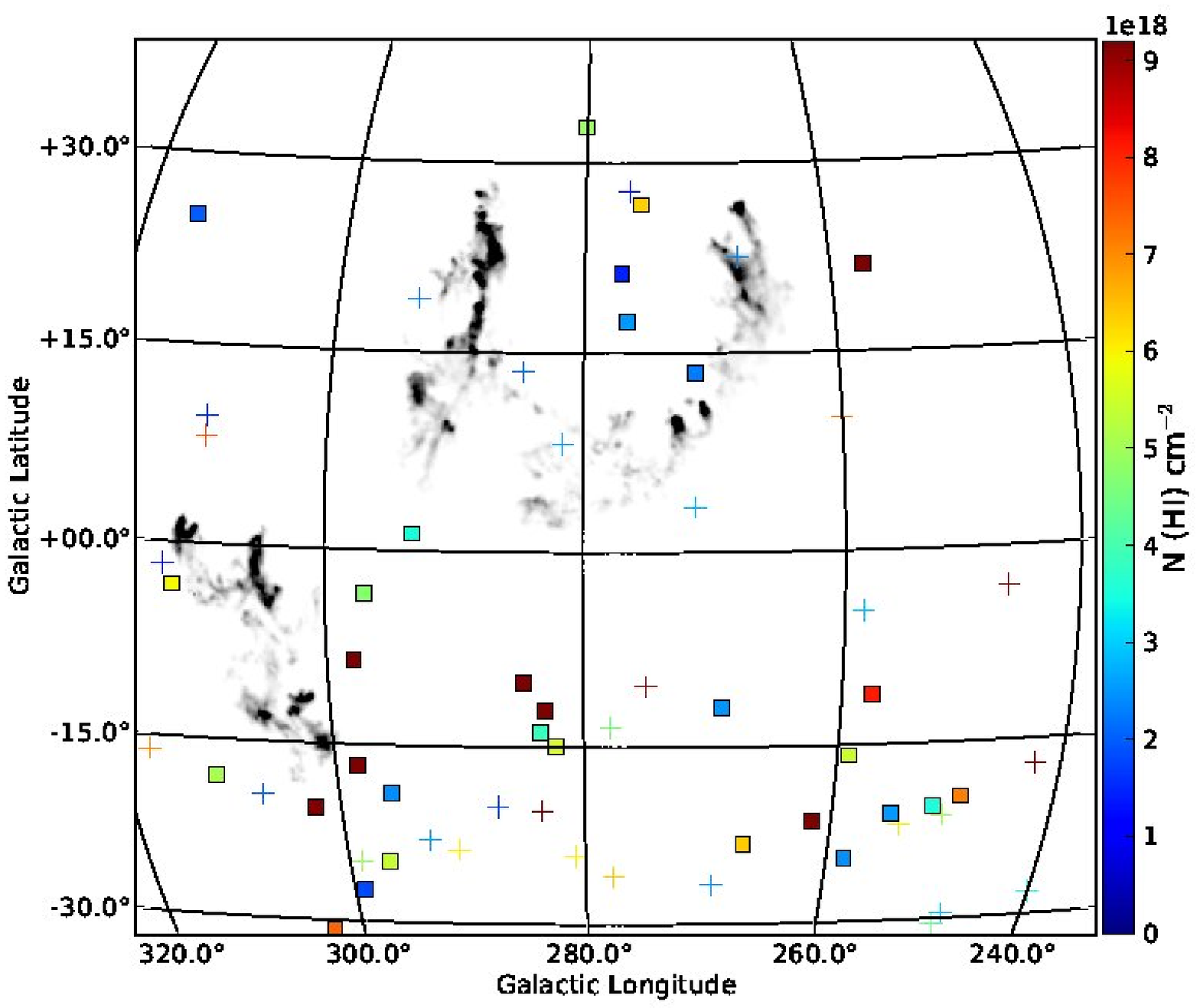}
\includegraphics[scale=0.45]{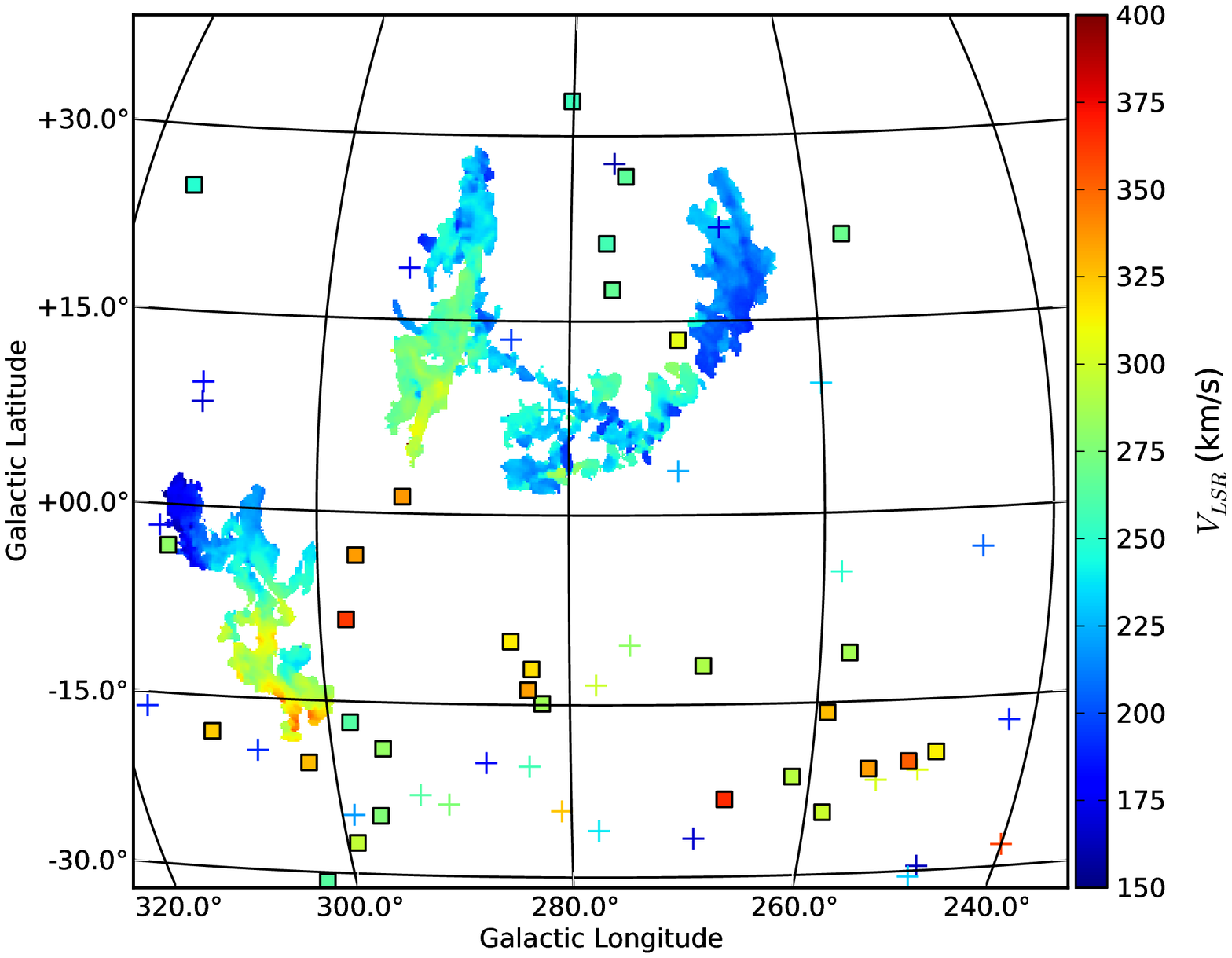}
\caption{Sky distribution of head-tail clouds with positive velocity gradient 
(pHT; pluses) and with negative velocity gradient (nHT; squares). 
See definition in \S\ref{HTclouds}. 
The top and bottom panels show the distributions in 
peak \HI\ column density and $V_{LSR}$, respectively.   
The colors indicate the values according to the color bar scale on the right side. 
\label{HTcomp}}
\end{center}
\end{figure}

To analyze the two subgroups of head-tail clouds with velocity gradient 
(\ie, pHT and nHT), we plot them with different symbols in Figure~\ref{HTcomp}. 
The plus and square symbols represent the pHT and nHT clouds, respectively. 
There are approximately equal numbers of pHT clouds and nHT clouds above and below the Galactic plane 
for both the peak \HI\ column density and $V_{LSR}$ distributions. 
As for the distribution of $V_{LSR}$ (bottom panel), pHT clouds possess slightly lower 
$V_{LSR}$ ($< 200$~\kms) than nHT clouds ($< 250$~\kms) above the Galactic plane. 
Below the Galactic plane, pHT clouds are evenly distributed across the range of $V_{LSR}$, 
and nHT clouds dominate at $V_{LSR} >250$~\kms. 
 
\subsection{Symmetric and Bow-Shock Clouds\label{SBclouds}}

The bow-shock shaped cloud is characterized by a dense core with two deflected 
gas wings, which have lower column density than the core. The presence of this 
type of cloud suggests ram-pressure interaction with the ambient medium. 
On the other hand, symmetric clouds do not exhibit any morphological signs of 
disturbance.  We note, however, that a head-tail cloud aligned with
the major axis along the line of sight would also appear as a symmetric cloud. 

\begin{figure}
\begin{center}
\includegraphics[scale=0.53]{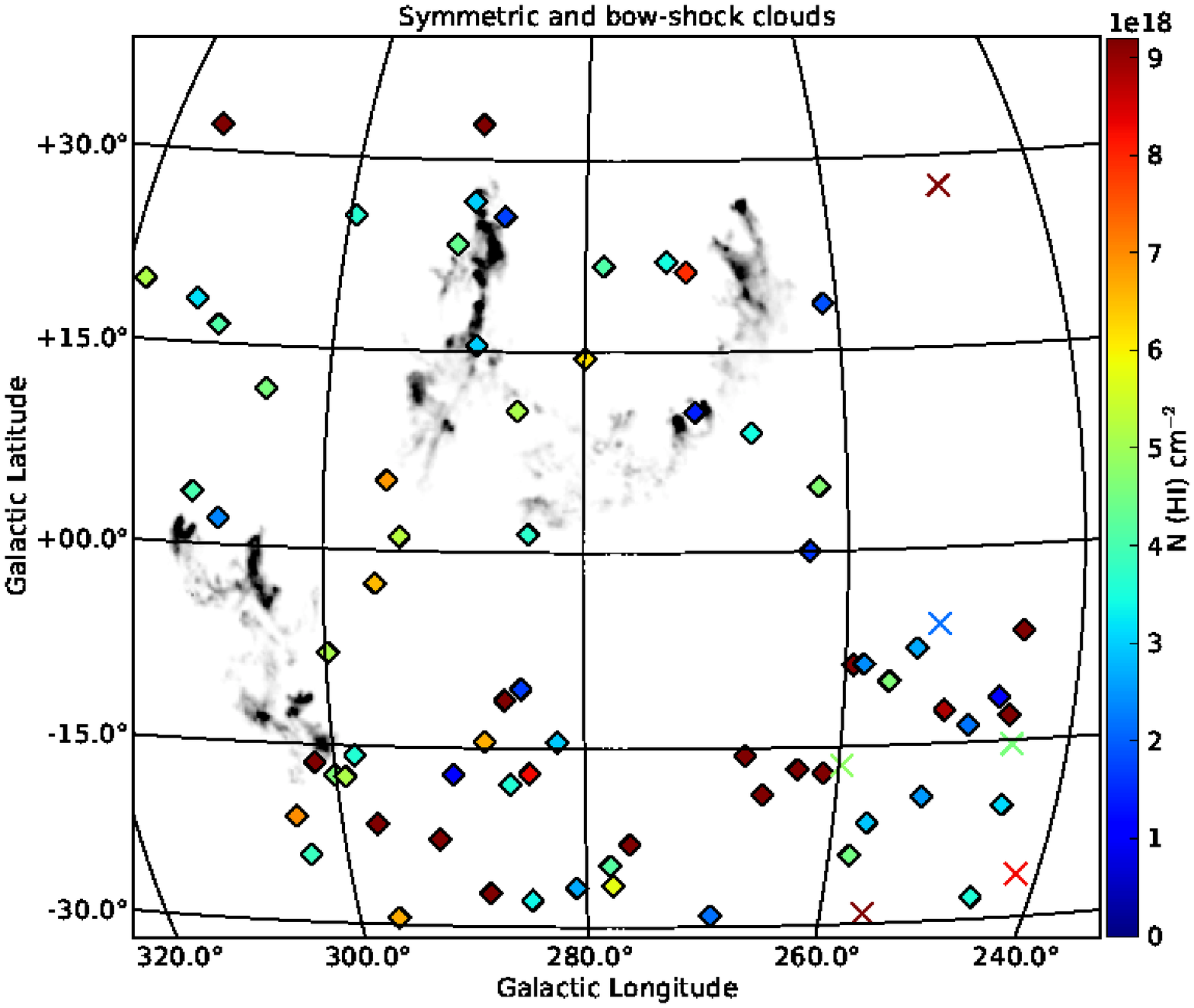}
\includegraphics[scale=0.55]{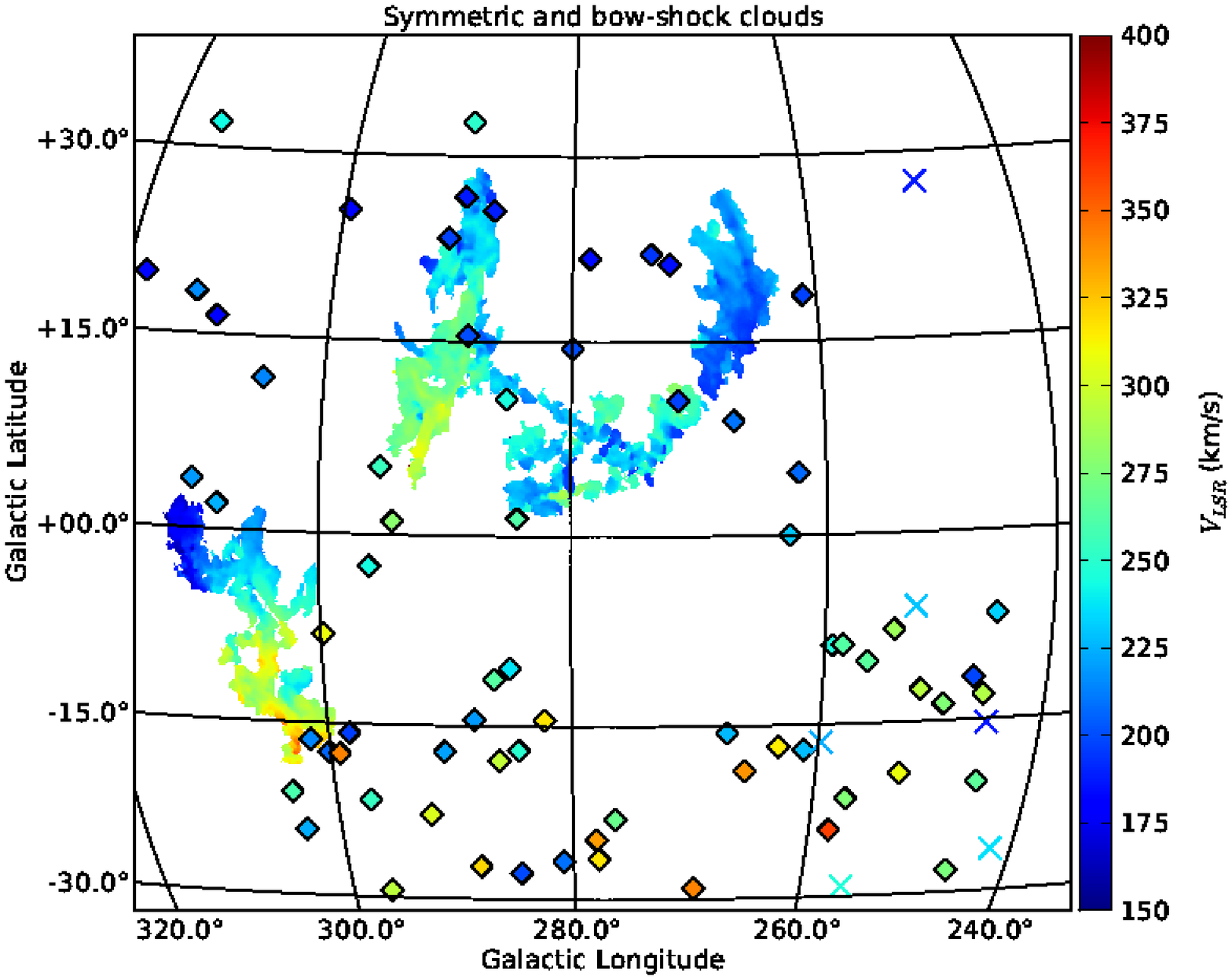}
\caption{Same as Figure~\ref{HTcomp}, except showing the sky distributions of symmetric (diamond) 
and bow-shock shaped clouds (crosses). 
\label{SB}}
\end{center}
\end{figure}

In Figure~\ref{SB}, we show the distributions of symmetric 
(diamonds) and bow-shock shaped (crosses) clouds in 
peak \HI\ column density and $V_{LSR}$. 
There are only few high \HI\ column density ($> 10^{18}$~cm$^{-2}$) 
symmetric clouds are found above the Galactic plane. 
Otherwise, they cover a wide range of $N_{\rm HI}$ in the region of the LA. 
The majority of these clouds fall between 225~\kms\ and 340~\kms\ 
below the Galactic plane and less than 250~\kms\ above the Galactic plane. 
With a small sample of bow-shock shaped clouds, we conclude that 
the typical bow-shock shaped cloud has $N_{\rm HI}\sim0.2-1\times 10^{19}$~cm$^{-2}$ 
and velocity less than 250~\kms. 

34$\%$ (23/69) of symmetric clouds also exhibit a velocity gradient. 
Velocity gradients in symmetric clouds can be caused by 
several effects.
For example, there may be 
an angle between the HVC velocity vector and the line of sight; 
there could be rotation in the clouds; 
two or more HVCs are superimposed on the same line of sight 
\citep{Bruns00}; and fluctuations in 
the background \HI\ emission \citep{Begum10}. 
There is no preferred direction of the velocity gradient for symmetric clouds in our catalog. 
We rule out projection of LSR velocity as the cause of 
velocity gradient. 
The typical velocity gradient is $\sim$0.3--0.8~\kms\ arcmin$^{-1}$, which is 
similar to the range (0.5--1 \kms\ arcmin$^{-1}$) 
found among the compact clouds studied by \citet{Begum10}. 
We note that our clouds are unresolved and thus subject to the effect of beam smearing. 

\section{Discussion\label{diss}}

\subsection{Implications of LA morphology for the origin of the LA}

\begin{figure}
\epsscale{1.0}
\plotone{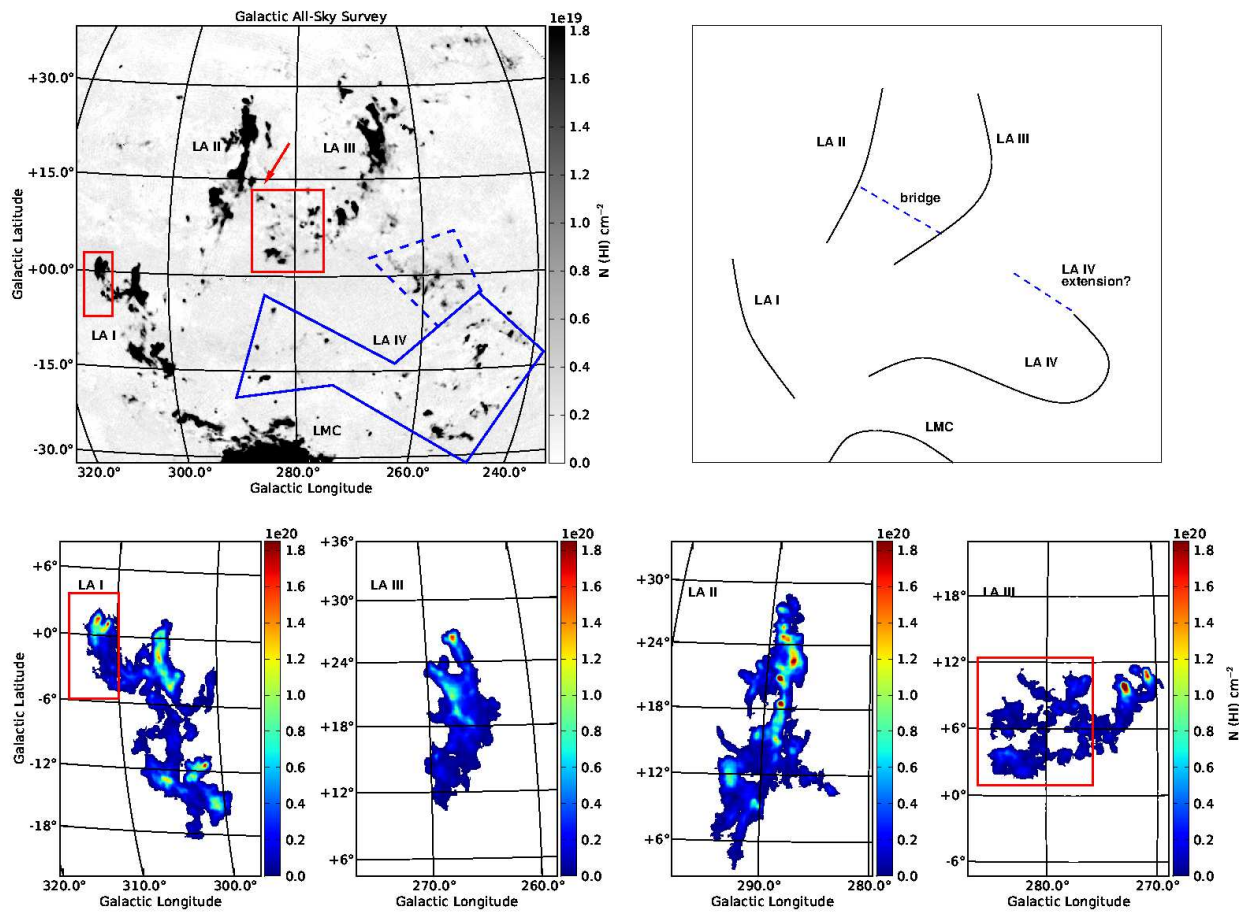}
\caption{The top left figure shows the integrated \HI\ column density map in the region of the Leading Arm. 
The red boxes highlight the extended features of LA complexes as seen in GASS. 
The arrow indicates the ``bridge'' structure connecting LA II and second cloud complex 
of LA III. The blue box shows the location of the new cloud population, namely LA IV. The blue dashed 
box marks the boundary of possible extended feature of the LA IV. 
The top right figure is a schematic diagram of the LA features and the LMC. 
The bottom subfigures show individual integrated \HI\ column density maps of LA complexes 
as identified by {\it Duchamp}. The red boxes match the position as shown in the top figure. 
\label{extension}}
\end{figure}

With the large sky coverage of GASS, we have uncovered new 
extended features of the Magellanic Leading Arm. 
In Figure~\ref{extension}, we show the relative position of 
the LA I, LA II and LA III (top figures) 
and individual LA complexes as identified by {\it Duchamp} (bottom subfigures).  
The red boxes highlight the extended features that were not detected in Br\"{u}ns' survey. 
The extended feature in LA I has been seen 
in other all-sky HVCs map (see \eg, \citealp{Putman02}).  
It was not detected in the Br\"{u}ns' survey due to its longitude coverage cutoff at 310\degr\ 
near the LA I region. 

The other extended feature seen in LA III is new. 
This feature consists of clumps connected by diffuse, 
low \HI\ column density filaments. The most interesting 
part about this extended feature is what appears to be a ``bridge'' connecting 
the LA II (see the arrow in top figure). The velocity map 
of this extended feature also shows similar velocity 
between LA II and LA III near the ``bridge''. 
This implies that LA II and LA III might have been part of 
a larger cloud complex in the past and have been pulled apart.

Here we report a new population of clouds, named LA IV, 
that is located South of the Galactic plane and to the north-west of the LMC. 
Although LA IV has also recently been reported by V12, 
the defined boundary of LA IV is more extended in this study than V12 
(see the schematic diagram of the LA IV feature at the top right panel of 
Figure~\ref{extension}). 
The median $V_{LSR}$ of LA IV is $\sim+260$~\kms.   
The blue and dashed lines mark the estimated boundary 
and extended boundary of the population 
in the top left panel of Figure~\ref{extension}, respectively. 
The morphology of LA IV is different from its counterparts, 
the LA I, II and III complexes. It is formed by a stream of cloudlets 
rather than a large complex, 
which the majority of them are head-tail and symmetric clouds, and spanning 
$\sim$50\degr\ across the sky.

The origin of the LA is somewhat controversial. The overall velocity structure 
of the \HI\ gas suggests that the LA originates from the LMC \citep{Putman98}. 
The coincidence of the LA I position and morphology with the LMC southeast \HI\ 
overdensity region further supports this observational evidence \citep{Nidever08}. 
While metallicity measurements from HST spectra of background sources 
toward the MS and LA II only constraint their origin to the MCs ($Z=0.2-0.4$~solar; 
\citealp{Lu94,Lu98,Gibson00}), new measurements of FUSE and HST spectra 
suggest that they originate from the SMC \citep{Sembach01,Fox10}. 
This observational evidence is supported by various simulations 
(see \citealp{Connors06,DB12}).

If we ignore the controversy and assume that the LA 
has a single origin, it does not explain why the LA IV has a 
different morphology 
than its counterparts. A possible explanation is that the LA IV has a 
different origin. In fact, the stream of cloudlets that form the LA IV 
appears to trace back to the LMC (see Figure~\ref{extension}). 
This explanation would fit in the model scenario of SMC origin 
for the LA I, LA II and LA III. 
Nevertheless, measurement of metallicity using background sources toward 
the LA IV and future simulations that incorporate the LA IV 
are necessary.

Distance must have an effect on the morphology because the \HI\ gas can 
interact with different ambient medium in different regions of the Galactic halo. 
With the exception of the LMC and SMC distances (50 and 60~kpc, respectively), 
distances to the LA and MS are hard to determine. 
According to the tidal models, the MS is further away from 
the Galactic plane than the MCs and the LA, 
with distances of 50--100~kpc at its tip \citep{YN03,Connors06}. 
An empirical study of filaments near the tip of the MS estimated 
a distance of 70~kpc, which is in agreement with the 
tidal models \citep{Stanimirovic08}. The LA is at an approximate kinematic distance 
of 21~kpc \citep{MG08}, based on the 
evidence of interaction between the LA I with the Galactic disk gas. 
This distance is smaller than the estimate due to 
cloud disruption timescale, in which the gas stream 
from the MCs is not expected to reach the disk 
in the form of \HI\ clouds \citep{HP09}. 
Simulations also put the distance 
of the LA to $\sim50$~kpc (see Figure~4 of \citealp{DB12}), although 
addition of ram-pressure stripping could result in a closer distance \citep{Connors06}. 

Examining the overall \HI\ gaseous feature of the Magellanic System, 
we find that the global cloud morphology 
in the LA region is strikingly similar to the northern extension region of the MS 
(\citealp{Nidever08,Stanimirovic08}). 
Both regions are rather clumpy and are populated with narrow line width clouds. 
These narrow line width clouds are generally exhibit multiphase structure, which 
consists of a cold core surrounded by a warm envelope. The existence of 
such multiphase \HI\ clouds along the northern tip of the MS, which is 
at a distance of 80~kpc, is rather surprising \citep{Stanimirovic08}.   
Future physical properties study of individual compact HVCs along the 
northern extension of the MS is needed to answer this question.

\subsection{Implications of head-tail clouds for the formation of the LA} 

The morphology of HVCs provides an important clue in studying the 
interaction between the neutral hydrogen gas and the ambient 
medium in the Galactic halo. 
Head-tail clouds are a classic example of cloud disruption via 
ram-pressure stripping when moving through the halo medium. 
They are relatively common as compared to the other morphological types. 
Such interaction commonly results in Kelvin-Holmholtz and 
thermal instabilities, and ultimately cloud fragmentation and evaporation 
\citep{Konz02}. 

Parameters such as the cloud size, halo and cloud densities have been shown 
to govern cloud stability in 3-dimensional hydrodynamical simulations
(see \eg, \citealp{HP09}; \citealp{QM01}; hereafter HP09 and QM01, respectively). 
In the QM01 models, pure gas and extragalactic dark-matter dominated HVCs 
with various gas densities, velocities and temperatures were investigated. 
They found that a tail with $N_{\rm HI} \geq 10^{19}$~cm$^{-2}$ appears when the 
external medium exceeds the density of 10$^{-4}$~cm$^{-3}$,  
although a weak, faint tail with $N_{\rm HI} \sim 10^{18}$~cm$^{-2}$ also 
becomes visible when the density of the external medium reaches $2\times10^{-5}$~cm$^{-3}$.  
The setup of the HP09 simulations was slightly different from QM01. 
They took into account the heating by an ultraviolet radiation field and 
metallicity-based cooling mechanisms. Various halo density profiles, 
cloud masses and velocities were tested for their wind-tunnel and free-fall models. 
The wind-tunnel model is best described as exposing the HVC to 
a wind with constant velocity and density. The free-fall model 
follows the trajectory of the HVC through an isothermal hydrostatic halo toward the disk. 
While both studies only consider clouds at lower $z$ (within 10~kpc for the HP09 models), 
and hence might not be suitable to explain the formation and evolution of HVCs originating from 
the Magellanic Clouds, it is quite interesting to note that the simulations 
have successfully simulated prominent head-tail clouds even with kinks or multiple cores 
that are reminiscent of the observational structure of head-tail clouds in this study. 
The simulated timescale for the cloud disruption strongly depends on the physical conditions
in the cloud and its interacting environment. Most head-tail clouds with high velocity are 
disrupted within 10~kpc and 100~Myr in the HP09 model but tail disruption can last as 
long as $\sim10^{9}$ yr in QM01 model after which the 
\HI\ column density drops below the observational threshold.  

The LA complexes are similar to head-tail clouds on a large scale. 
The directionality of their head and tail suggests that they are moving toward higher 
Galactic latitudes, although the curvature of LA II and LA III  
are in the opposite direction to LA I.
The velocity gradient shows that the velocity at the head is slower than 
the tail in LA I and II (see Figure~\ref{mom1_vlsr}).  
This kind of velocity gradient is expected when the head of the cloud is decelerated while   
moving through the ambient medium.  

Assuming that all small HVCs in the region of the LA (excluding 
cloudlets that form part of the LA IV) are fragments 
from the LA complexes due to cloud disruption, the directionality of the 
head-tail clouds should follow the direction of motion of the LA.  
However, this is not what we observed either for the head-tail clouds with or 
without velocity gradient (as mentioned in \S\ref{HTclouds}; see also 
Figure~\ref{HT_NHI}). The pointing direction of the head-tail clouds is random 
which suggests that turbulence in the medium must be at play. 
According to \citet{AH05}, this scenario can be generated when the 
incoming warm neutral gas collides with the 
hotter ambient medium and creates a thermally unstable region. 
If the flow is weakly turbulent, part of the warm gas condenses into cold gas, 
and any thermally unstable cold gas 
will continue to fragment until thermal equilibrium is achieved. 
In the case of a strongly turbulent incoming flow, 
which is the case for the interaction 
between the LA and the Milky Way halo medium (based on the velocity of the gas stream), 
the fragmented clouds should appear distorted and irregular. 
Strong turbulence also promotes the occurrence of fragmentation 
for the thermally unstable clouds, and subsequently, more small, cold clouds 
with lower density. This scenario appears to agree with the  
properties of HVCs in the region of the LA. 

In \S\ref{HTclouds}, we presented the dichotomy in radial velocity for the 
head-tail and symmetric HVCs above and below the Galactic plane 
(see Figures~\ref{HT_gradvel} and \ref{SB}), in which the HVCs below 
the Galactic plane possess lower $V_{LSR}$ than above the Galactic plane. 
Such dichotomy (or gradient as a function of Galactic latitude) is 
also seen in simulations,  
although the gradient has an offset between the model and 
the observational data (see Figure~7 of \citealp{DB12}). 
Simulations only include gravitational force, 
we suggest that an orbital effect is the cause of such dichotomy. 
Since we are only interested in 
the real distortion due to interaction of the cloud with the ambient medium, 
we will only consider the head-tail clouds with velocity gradient for 
the rest of this discussion. 
   
If we assume that all LA HVCs are traveling along the same orbit as
the MC's, we can use knowledge of the MC's total velocity and vector
components to estimate the tangential velocity ($V_{t}$) of the HVCs from
their measured radial velocities in the GSR frame.  Hence, we can
translate the median $V_{GSR}$ of HT clouds above and below the
Galactic plane into their associated tangential velocities ($V_{t}$)
based on the relative fractions of $V_{t}$ and $V_{GSR}$ of the MCs (LMC: $V_{t}=367$~\kms,
$V_{GSR}=89$~\kms; SMC: $V_{t}=301$~\kms, $V_{GSR}=23$~\kms;
\citealp{K06b}).  With the median $V_{GSR}$ of $-7$ and 54~\kms\ for
clouds above and below the Galactic plane, the estimated tangential
velocities are 270~\kms\ and 330~\kms, respectively.  
While the $V_{t}$ difference is small, 
we can estimate the mean density of the halo
region ($n_{h} (z)$) that the clouds move through,
\begin{equation}
C_{D}f_{c}n_{h}(z)=\frac{2N_{HI}g(z)}{v^2}
\end{equation}
\citep{BD97}, where $C_{D}$ is the drag coefficient, $f_{c}=N_{\rm HI}/(N_{\rm HI}+N_{\rm H II})$ 
is cloud neutral fraction, $g(z)$ is the gravitational acceleration, $N_{\rm HI}$ is 
the total \HI\ column density of the cloud and 
the $v$ is velocity of cloud.  
We assume that the total \HI\ column density of a typical head-tail cloud is 10$^{19}$~cm$^{-2}$, the 
drag coefficient and cloud neutral fraction are 1.0, and  
the gravitational acceleration is constant ($0.2\times10^{-8}$~cm s$^{-2}$) 
beyond 10~kpc \citep{Wolfire95} for determining the mean halo densities. 
For the velocities above and below the Galactic plane of 270~\kms\ and 330~\kms, 
we obtain $n_{h}$ = $5.4\times10^{-5}$ and $3.7\times10^{-5}$~cm$^{-3}$, respectively.     
We note that the calculation is sensitive to the adopted 
cloud \HI\ column density and uncertainty on $f_{c}$. Nevertheless, 
while the halo density is poorly known beyond 10~kpc, this exercise demonstrates that 
a slight difference in the halo density might 
cause the morphological differences between the LA I and LA II+LA III.  

The presence of a large number of head-tail clouds in the region of the 
LA as compared to the MS is somewhat curious (see PSM11). High angular 
resolution study of filaments near the northern tip of the MS has also shown the dominance of 
spherical clouds and an absence of elongated or head-tail clouds \citep{Stanimirovic08}. 
Assuming an isothermal halo, $P_{\rm ram}\propto\rho v^{2}$, where $\rho$ is 
halo density and is proportional to $d^{-2}$, the distances predicted from the \citet{DB12} 
simulations would suggest a higher ram-pressure interaction for the LA than the bulk of MS. 
Both observational evidence and theory support ram-pressure stripping as 
a more important factor than gravitational force for producing the morphological features of the LA. 
The inclusion of ram-pressure stripping in models 
could therefore result in closer predicted distances for the LA clouds, as suggested 
by the observational data. 
Finally, with the close distance of the LA, we suggest that fragmentation is not the only 
mechanism that produces the small 
clouds in the region. The cloud fragments with lower $z$-distances  
would experience increased background pressure which would increase the cooling rate. 
This would result in reforming of cold \HI\ clouds, as seen under the free-fall model of HP09.
 
\section{Summary and Conclusions\label{sc}}

We have produced a catalog of high velocity clouds in the region of
the Magellanic Leading Arm from Parkes Galactic All-Sky Survey data,
using the cloud search algorithm {\it Duchamp}.  We used {\it Duchamp}
to parametrize cloud properties including position, velocity 
and velocity FWHM.  We determined the angular size of
sources via 2-dimensional Gaussian fitting and peak \HI\ column
density via searching the brightest pixel in the integrated maps.
Comparison between our HVC catalog with that of \citet{Putman02} in
the same region and velocity range shows that the high spectral
resolution of GASS allows us to recover clouds with narrow line widths.
The total number of detected HVCs is 448 for P02 catalog and 419 for
our catalog. The combined catalog contains $\sim$625 unique clouds. 

We have presented the general distribution of HVCs in the catalog. 
The kinematic distributions with respect to Galactic longitude and latitude are 
generally consistent with the findings in P02. A trend of decreasing number of clouds 
from higher to lower Galactic latitude as velocity increases in all velocity 
reference frames was found. A morphological classification of 
clouds was presented, and distributions of each type 
were discussed. 

An extended feature in the LA I complex that was not covered 
in the detailed study of the Magellanic System by \citet{Bruns05} was noted. 
A new population of clouds that forms the LA IV and an extended feature that forms 
a diffuse ``bridge'' connecting the LA II and III complexes were also discovered.   
The discovery of the LA III extended feature demonstrates the importance of 
brightness temperature sensitivity and 
spectral resolution for an all-sky survey. Simulations have yet to 
reproduce this feature of LA. 

The most significant result in this study 
was the detection of a large number of head-tail clouds in the 
region of the LA as compared to the MS, suggesting that ram-pressure stripping is relatively 
more important than gravitational forces for the morphology and formation of the LA. 
The LA I and II themselves are large head-tail clouds, which are moving toward higher 
Galactic latitudes and both show a large velocity gradient, with the head 
being lower than the tail. We found that there was no preferred pointing 
direction for the small head-tail clouds. This suggests a scenario
where the clouds are produced in a turbulent flow where incoming 
warm neutral gas collides with the hot halo ISM. 
The cloud morphologies are strongly correlated to the degree of turbulence in the ISM. 
The presence of strong turbulence is probably the cause for the observed 
morphologies and properties of clouds in the region. 

A dichotomy in velocity for the head-tail and symmetric HVCs above and
below the Galactic plane was found. Since such dichotomy is also seen 
in simulations of \citet{DB12}, we suggest that an orbital effect is the cause. 
Finally, using the typical \HI\
column density of cloud and tangential velocities above and below the
Galactic plane, we infer a small difference in halo density. 
This suggests that the LA II and LA II+LA III are interacting with different halo 
environments, which might explain the morphological difference between them.

\appendix

\section{APPENDIX: COMPARISON WITH V12 CATALOG}

\begin{figure}
\plotone{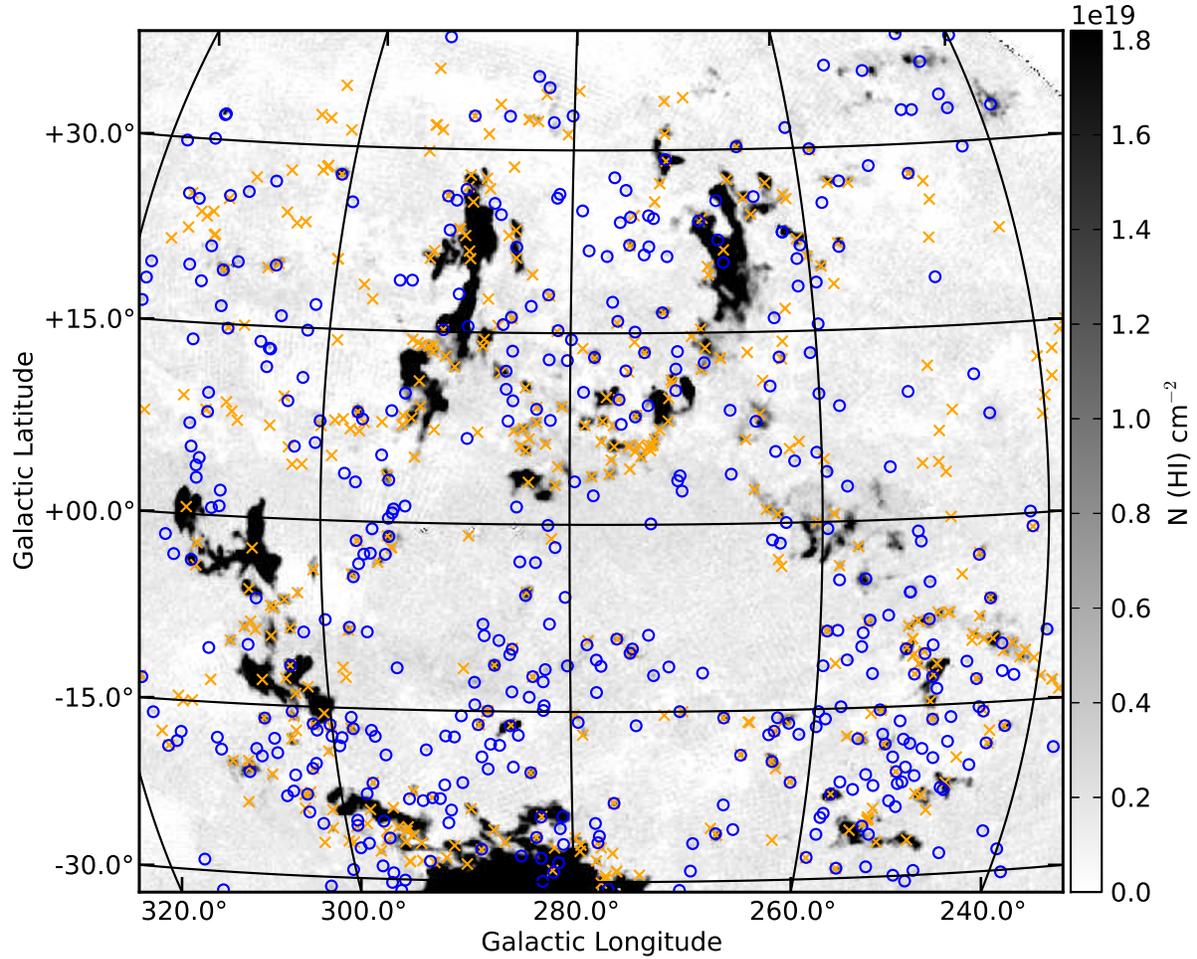}
\caption{On-sky distribution of the 431 sources detected by {\it Duchamp} (blue circles) 
and the 433 sources detected by \citet{Venzmer12} (orange crosses) 
in the same studied region of the Leading Arm. 
The integrated \HI\ column density map of Figure~\ref{LA_mom0} is shown. 
There is a slight difference in the studied velocity range of two catalogs. 
\label{duchampvsimagej}}
\end{figure}

While this paper was under review, a similar study appeared in press by \citet{Venzmer12}. 
Here we compare our catalog to the V12 catalog, which was compiled using  
a similar studied area to this paper. \citet{Venzmer12} extracted the GASS 
data cube from 52~\kms\ to 400~\kms\ $V_{LSR}$ and performed the source 
finding and parametrization with the image processing software, $ImageJ$. 
We refer the reader to V12 for a detailed description of source selection criteria. 
In Figure~\ref{duchampvsimagej}, we show a comparison between HVCs from the V12 catalog 
that fall within our studied area. 
The total identified HVCs in this region is 419 (our catalog) vs 433 (V12). 
Sixteen of HVCs in V12 catalog lie outside our studied area. 

While the total number of clouds between the two catalog is similar, only 
$\sim$120 HVCs are identified to be the same cloud in the two catalogs. 
The differences are due to: 
(a) degrees of breaking up or merging clouds in extended complexes; 
(b) the searched velocity range; and 
(c) selection criteria for clouds. 
The degrees of breaking up or merging clouds can be easily seen in 
Figure~\ref{duchampvsimagej} (\eg, near the region of LA~I), which 
$ImageJ$ breaks up more clouds than $Duchamp$ for a given region. 
There are pros and cons for both source finding softwares in term of degrees 
of breaking up or merging clouds. The advantage of breaking up a large 
cloud complex into smaller sub-clouds allows analysis of individual sub-clouds 
(\eg, breaking up the LA I into LA 1.1--1.3). However, 
a high degree of breaking-up of clouds can result in missed detections 
such as the diffuse bridge-like feature we see connecting LA II and LA III.

\begin{figure}
\plotone{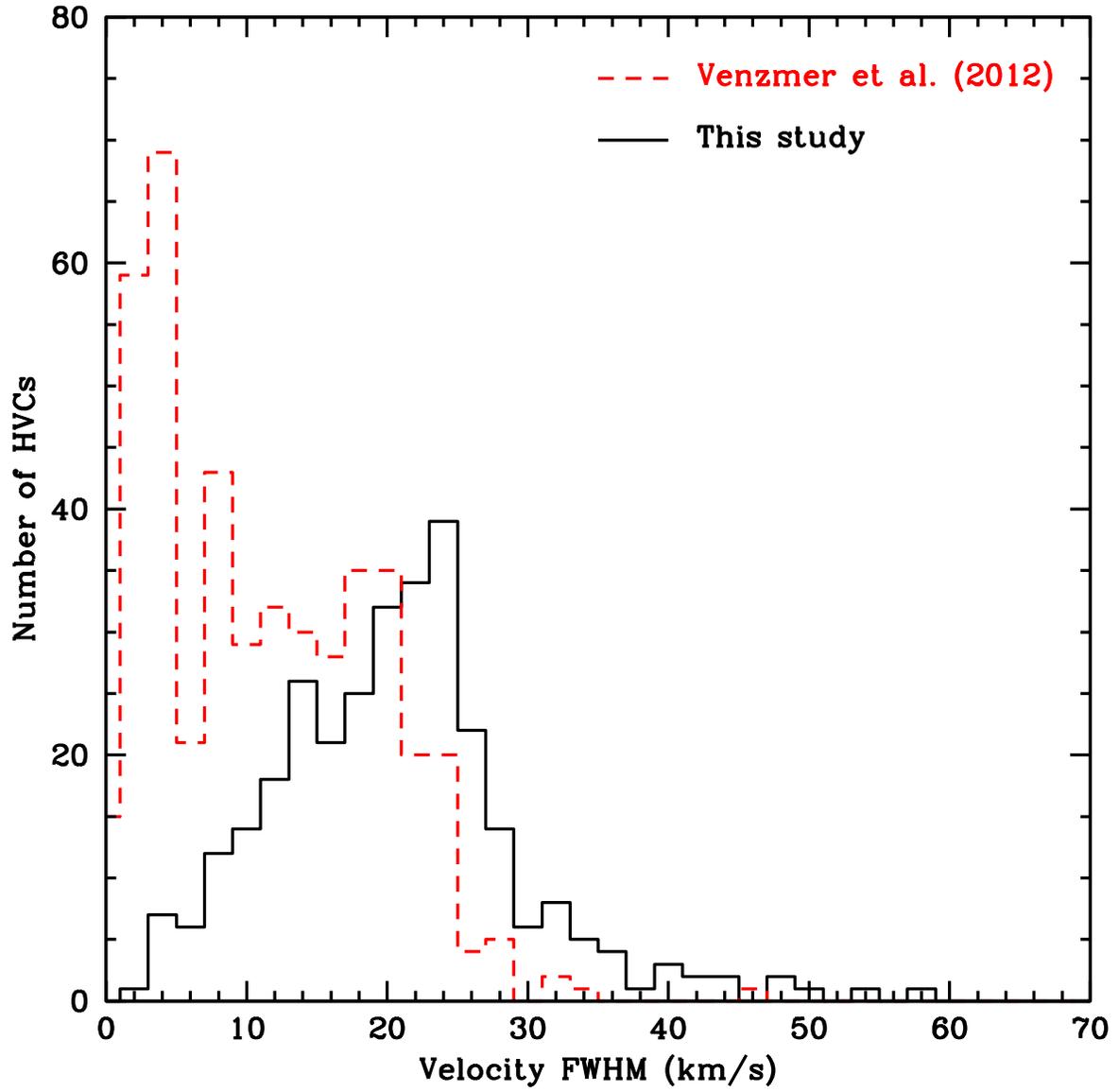}
\caption{
Similar to the bottom panel of Figure~\ref{hist_HTcolvel}: 
histograms of velocity FWHM of HVCs in V12 (dashed line) and this study (solid line). 
\label{wvel_duchampvsimagej}}
\end{figure}

As mentioned in \S 2, the Galactic \HI\ emission is extremely strong in the velocity range 
0~\kms~$\leq$~$V_{LSR}$~$<$~150~\kms. Mild contamination is also 
evident between 150 and 190~\kms. Part of the searched velocity range in V12 
and this study are affected by Galactic \HI\ emission. While we manually 
masked out the affected region, V12 applied a cut off to the velocity FWHM 
that excludes the Galactic \HI\ emission. 
A comparison of velocity FWHM distributions in both catalogs 
in Figure~\ref{wvel_duchampvsimagej} reveals a poor correlation. 
About $50\%$ of clouds in V12 catalog have narrow line widths (defined here as velocity FWHM 
less than 10~\kms). Close examination of a subset of the V12 narrow line width HVCs 
shows that some are real detection of small clumps, but some are false detections 
due to artifacts and noise peaks. Furthermore, there are a significant number of clouds 
with quoted velocity width much lower than we measure - in some cases as low as the GASS 
velocity resolution.

\begin{figure}
\plotone{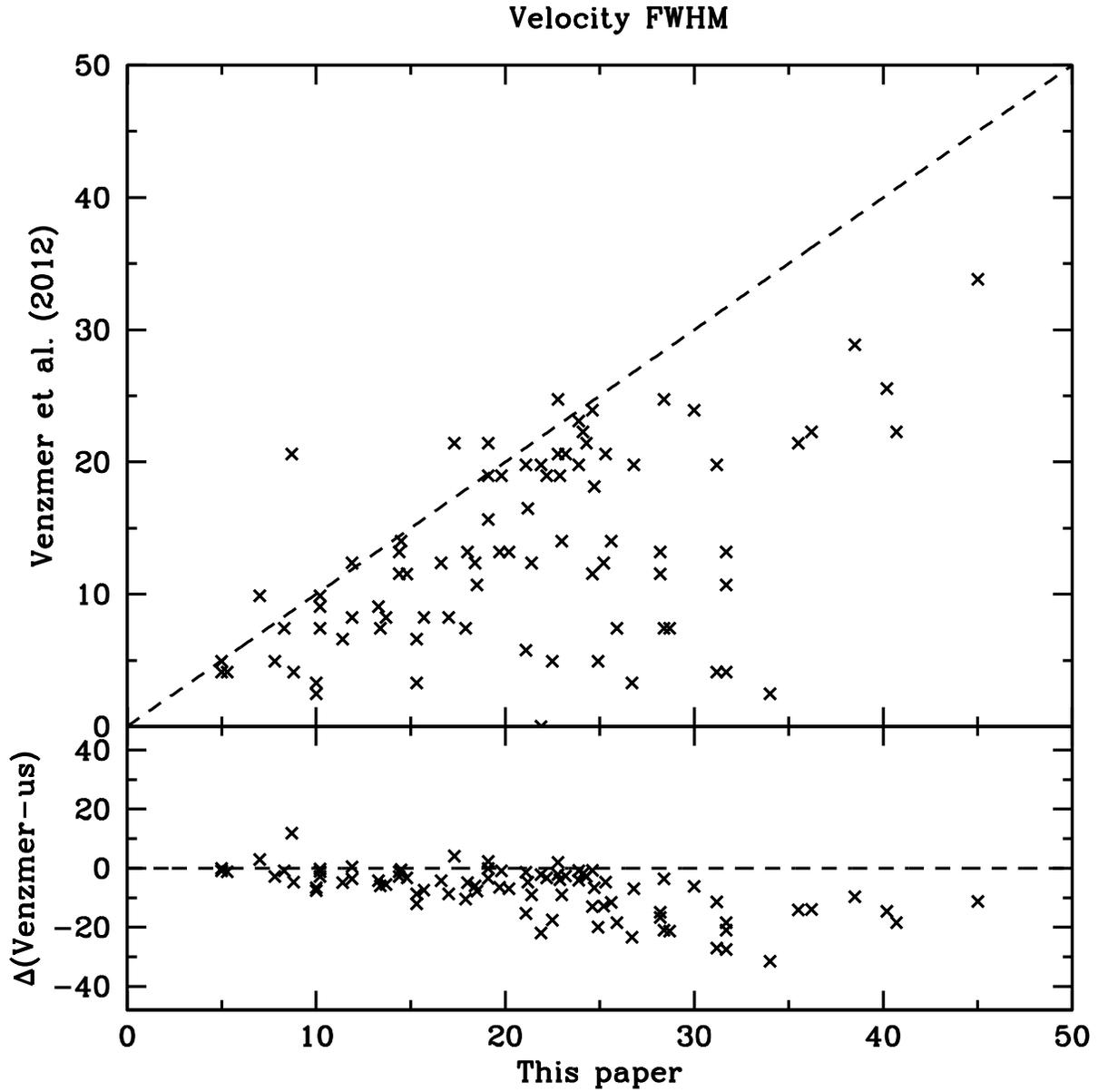}
\caption{
Comparison of our measured velocity FWHM with V12. The top panel shows 1:1 comparison of velocity FWHM measurements. 
The bottom panel shows the difference between our measurements and V12. One HVC with velocity FWHM larger than 
50~\kms\ is not shown here. 
\label{dvsv12_fwhm}}
\end{figure}

\begin{figure}
\plotone{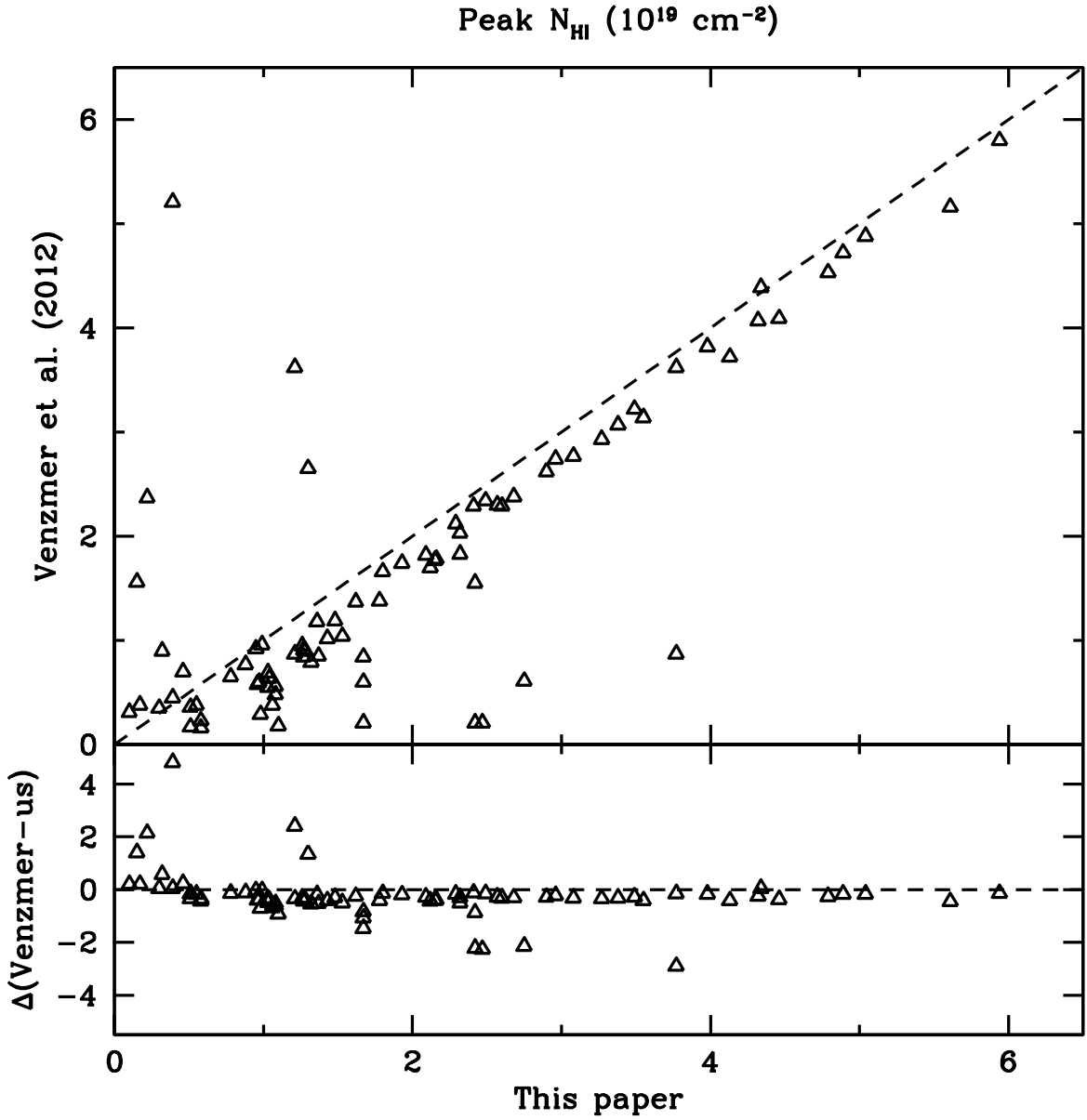}
\caption{
Same as Figure~\ref{dvsv12_fwhm} except comparing peak $N_{\rm HI}$. One HVC with peak $N_{\rm HI}$ larger than $6\times10^{19}$~cm$^{-2}$ 
is not shown here. 
\label{dvsv12_nHI}}
\end{figure}

We compare the velocity FWHM and peak $N_{\rm HI}$ values of HVCs that are found in both catalogs. 
HVCs identified as the same source in the two catalogs, but without listed parameters in our catalog, 
are excluded from the comparison.
Figures~\ref{dvsv12_fwhm} and \ref{dvsv12_nHI} show the comparison of 
the measurements (top panels) 
and the difference in measurements (bottom panels). 
We find $<$$\Delta$V$_{\rm FWHM}$$>=-8.3$~\kms, $\sigma=9.4$~\kms, 85 HVCs;
the velocity FWHM measured in V12 is systematically lower than ours. 
The difference is more significant for broad line width HVCs ($>$20~\kms). 
Direct comparison for the velocity FWHM of other small 
clumps as detected in V12 cannot be made because 
they have been merged into larger clouds by $Duchamp$. Nevertheless, 
examining some of these HVCs in the GASS data cube, we find that the measurements in V12
are consistently underestimated. For peak $N_{\rm HI}$, we find 
 $<\Delta$N$_{\rm HI}>=-0.25\times10^{19}$~cm$^{-2}$, $\sigma=0.91\times10^{19}$~\kms, 85 HVCs. 
The peak $N_{\rm HI}$ measurements in V12 is also systematically lower than ours 
and with large differences for some of the HVCs. 
 
\acknowledgments

BQF is the recipient of a John Stocker Postdoctoral Fellowship from the 
Science and Industry Research Fund. This research made use of APLpy, 
an open-source plotting package for Python hosted at http://aplpy.github.com 
and data from the Parkes Galactic All-Sky Survey. The Parkes radio telescope 
is part of the Australia Telescope National Facility which is funded by the 
Commonwealth of Australia for operation as a National Facility managed by CSIRO. 
We thank Tobias Westmeier for providing his parametrization algorithm and helpful 
comments; Matthew Whiting for providing helps on running {\it Duchamp}; Kenji Bekki 
for generating a fruitful discussion and the anonymous referee for his/her comments 
to help improving the original manuscript.   

\bibliographystyle{apj}
\bibliography{ref}

\clearpage

\clearpage

\clearpage

\clearpage

\clearpage

\clearpage

\clearpage

\clearpage

\clearpage

\clearpage

\begin{deluxetable}{cccccccccccccc}
\rotate
\tabletypesize{\tiny}
\tablewidth{0pt}
\tablecolumns{14}
\tablecaption{GASS catalog of detected sources in the region 
of the Magellanic Leading Arm.\label{catalog}}
\tablehead{
\colhead{ID} & 
\colhead{Designation} &
\colhead{$V_{LSR}$} &
\colhead{$V_{GSR}$} &
\colhead{$V_{LGSR}$} &
\colhead{FWHM} &
\colhead{$F_{\rm int}$\tablenotemark{a}} &
\colhead{$T_{\rm B}$} &
\colhead{$N_{\rm HI}$}  &
\colhead{Semi-major}  & 
\colhead{Semi-minor}  & 
\colhead{PA}  &
\colhead{Flag\tablenotemark{b}} & 
\colhead{Classification\tablenotemark{c}}  \\
\colhead{} & 
\colhead{($gl \pm gb +V_{LSR}$)} &
\colhead{\kms} &
\colhead{\kms} &
\colhead{\kms} &
\colhead{\kms} &
\colhead{Jy km s$^{-1}$} &
\colhead{K} &
\colhead{10$^{19}$ cm$^{-2}$}  &
\colhead{$\degr$}  & 
\colhead{$\degr$}  & 
\colhead{$\degr$}  &
\colhead{} & 
\colhead{}  \\
\colhead{(1)} & 
\colhead{(2)} &
\colhead{(3)} &
\colhead{(4)} &
\colhead{(5)} &
\colhead{(6)} &
\colhead{(7)} &
\colhead{(8)}  &
\colhead{(9)}  & 
\colhead{(10)}  & 
\colhead{(11)}  &
\colhead{(12)} & 
\colhead{(13)} &
\colhead{(14)} \\
}
\startdata
63	&	HVC+236.9$-$19.1+174	&	\nodata	&	\nodata	&	\nodata	&  \nodata	&     	\nodata	&	\nodata	&	\nodata	&	\nodata	&	\nodata	&	\nodata	&	M	&	IC	\\
64	&	HVC+307.2+26.8+174	&	174.0	&	17.5	&	$-$60.2	&  12.4	&     	2.1	&	0.20	&	0.23	&	0.2	&	0.2	&	$-$54	&	\nodata	&	:HT	\\
65	&	HVC+273.5+24.5+174	&	174.2	&	$-$25.6	&	$-$79.9	&  4.3	&     	1.3	&	0.17	&	0.09	&	0.3	&	0.2	&	54	&	\nodata	&	IC	\\
66	&	HVC+251.3+36.6+175	&	174.5	&	7.3	&	$-$28.0	&  48.0	&     	174.8	&	0.62	&	2.63	&	\nodata	&	\nodata	&	\nodata	&	\nodata	&	IC	\\
67	&	HVC+293.3+19.1+175	&	174.8	&	$-$16.2	&	$-$85.5	&  11.5	&     	3.1	&	0.20	&	0.22	&	0.3	&	0.2	&	35	&	\nodata	&	HT	\\
68	&	HVC+313.9$-$01.7+175	&	175.2	&	16.8	&	$-$54.0	&  8.7	&     	1.5	&	0.21	&	0.12	&	0.3	&	0.2	&	$-$37	&	\nodata	&	HT	
\enddata
\tablecomments{Table~\ref{catalog} is published in its entirety in the
electronic edition of the {\it Astrophysical Journal}.  
A portion is shown here for guidance regarding its form and content.}
\tablenotetext{a}{Corrected $F_{\rm int}$.}
\tablenotetext{b}{SR: the detection lies at the edge of the spectral region; 
M: the detection extends over to the masked Milky Way emission region; 
E: the detection is next to the spatial edge of the image.}
\tablenotetext{c}{HT: head-tail cloud with velocity gradient; :HT: head-tail 
cloud without velocity gradient; S: symmetric cloud; B: bow-shock cloud; 
IC: irregular/complex cloud.}
\end{deluxetable}

\end{document}